\RequirePackage{silence}
\documentclass{aa}
\AtBeginDocument{\let\linenumbers\relax}

\usepackage{graphicx}
\usepackage{amsmath,amssymb}
\usepackage{booktabs}
\usepackage{xcolor}
\usepackage{txfonts}
\usepackage{placeins}
\usepackage{xspace}

\usepackage{silence}
\usepackage[breaklinks,bookmarks=false,pdfpagemode=UseNone]{hyperref}

\graphicspath{{figures/}}
\usepackage{physics}

\usepackage{dcolumn}
    \newcolumntype{d}[1]{D{.}{\cdot}{#1}}
    \newcolumntype{.}{D{.}{.}{-2}}
    \newcolumntype{,}{D{,}{,}{-3}}

\newbox\tablebox    \newdimen\tablewidth \def\leaderfil{\leaders\hbox to 5pt{\hss.\hss}\hfil}

\usepackage{xspace}

\usepackage{tikz}
\usetikzlibrary{positioning, fit, calc, arrows.meta, backgrounds}

\definecolor{CFblue}{RGB}{60, 100, 165}
\definecolor{CForange}{RGB}{210, 135, 35}
\definecolor{CFgreen}{RGB}{75, 140, 80}

\usepackage[breaklinks,bookmarks=false,pdfpagemode=UseNone]{hyperref}

\usepackage{fontawesome5}
\providecommand{\orcid}[1]{\,\href{https://orcid.org/#1}{\textcolor[HTML]{A6CE39}{\faOrcid}}}

\newcommand{\qube}{\texttt{QUBE}\xspace}
\newcommand{\cosmoforge}{\texttt{CosmoForge}\xspace}
\newcommand{\pics}{\texttt{PICSLike}\xspace}

\newcommand{\dvec}{\mathbf{d}}            
\newcommand{\svec}{\mathbf{s}}            
\newcommand{\nvec}{\mathbf{n}}            
\newcommand{\xvec}{\mathbf{x}}            
\newcommand{\yvec}{\mathbf{y}}            
\newcommand{\Cmat}{\mathbf{C}}            
\newcommand{\Smat}{\mathbf{S}}            
\newcommand{\Nmat}{\mathbf{N}}            
\newcommand{\Imat}{\mathbf{I}}            
\newcommand{\Pmat}{\mathbf{P}}            
\newcommand{\Fmat}{\mathbf{F}}            
\newcommand{\Rmat}{\mathbf{R}}            
\newcommand{\Umat}{\mathbf{U}}            
\newcommand{\Zmat}{\mathbf{Z}}            
\newcommand{\Dmat}{\mathbf{D}}            
\newcommand{\Omat}{\mathbf{O}}            
\newcommand{\Lchol}{\mathbf{L}}            
\newcommand{\Proj}{\boldsymbol{\Pi}}       
\newcommand{\Tmat}{\mathbf{T}}            
\newcommand{\tp}{^{\mathsf T}}            

\newcommand{\pcl}{pseudo-$C_\ell$\xspace}
\newcommand{\Cl}{C_\ell}
\newcommand{\sigr}{\sigma(r)}
\newcommand{\sigA}{\sigma_A}

\newcommand{\fsky}{f_{\rm sky}}
\newcommand{\ukam}{\,\mu\mathrm{K\,arcmin}}

\begin{document}

\title{CosmoForge}

\subtitle{II. An exact treatment of arbitrary linear filters in quadratic maximum-likelihood power spectrum estimation}

\titlerunning{CosmoForge. II}
\authorrunning{Galloni}

\author{
G. Galloni\inst{1,2}\corrauth{giacomo.galloni@unife.it}
}

\institute{
Dipartimento di Fisica e Scienze della Terra, Universit\`a degli Studi di Ferrara, via Saragat 1, I-44122 Ferrara, Italy
\and Istituto Nazionale di Fisica Nucleare, Sezione di Ferrara, via Saragat 1, I-44122 Ferrara, Italy
}
\date{Received / Accepted}

\abstract
{Ground-based cosmic microwave background (CMB) experiments filter their time-ordered data to remove $1/f$ noise and ground pickup, which removes or mixes the large angular scales where primordial $B$ modes peak. A common remedy is a \pcl estimator with harmonic $B$-purification and a transfer function measured from simulations. Another applies it after a matrix-based purification that separates $E$ from $B$ modes in the maps.}
{We extend the quadratic maximum-likelihood (QML) estimator of \cosmoforge to filtered maps, treating the filter exactly rather than correcting for it afterwards, and compare its error bars with both \pcl approaches under several map-level filters.}
{A linear filter enters the estimator by conjugating its ingredients, without altering its structure. The same holds for the pixel-based likelihood of \cosmoforge, which gives the smallest error bar achievable on the same maps. We compare all methods in a simplified setup through the error bars on bandpowers and on the tensor-to-scalar ratio $r$.}
{QML sits $2$ to $4\%$ above the pixel-likelihood bound on $\sigr$ in every filter state. At low noise, the pure \pcl $\sigr$ is $1.21$ times QML's without filtering, $3.0$ times under a scan filter along stripes of constant declination, and $3.7$ times under one from a toy constant-elevation scan. The matrix-purified \pcl $\sigr$ is at most $1.03$ times QML's except under the constant-elevation filters, where it reaches $1.10$. These factors shrink towards unity at higher noise.}
{The pure \pcl gap comes from apodisation and from harmonic $E/B$ separation after the filter, and pixel-space separation closes it, except under the constant-elevation filters, which attenuate modes rather than remove them and widen both gaps. QML stays within a few per cent of the optimal error bars without the tuning and the simulated transfer functions of the \pcl pipelines. Thus, \cosmoforge provides a robust framework for power-spectrum estimation and inference on filtered maps at degree scales.}


\maketitle

\section{Introduction}
\label{sec:intro}

A detection of primordial $B$-mode polarisation in the cosmic microwave background (CMB) \citep{KamionkowskiKosowskyStebbins1997, SeljakZaldarriaga1997} would measure the tensor-to-scalar ratio $r$, i.e. the amplitude of primordial gravitational waves in units of scalar perturbations, and through it, the energy scale of inflation \citep{KamionkowskiKovetz2016}. Indeed, the linear polarisation of the CMB splits into a gradient-like $E$ and a curl-like $B$ component, and at linear order primordial scalar perturbations produce only the first \citep{KamionkowskiKosowskyStebbins1997, SeljakZaldarriaga1997}. We write $EE$ and $BB$ for their angular power spectra. Satellite missions fixed the $\Lambda$ cold dark matter ($\Lambda$CDM) background against which a tensor component must be distinguished: the Wilkinson Microwave Anisotropy Probe (WMAP) and \textit{Planck} measured the temperature anisotropies to the cosmic-variance limit on scales larger than about ten arcminutes, and mapped the $E$ modes over the full sky \citep{WMAP9, Planck2018VI}.

However, they did not detect primordial $B$ modes. The degree-scale searches happen from the ground, surveying one to ten per cent of the sky, such as BICEP/Keck and the Simons Observatory (SO) small aperture telescopes (SATs) \citep{BK18, SOforecast}. At these scales, the primordial signal must also be separated from the $B$ modes that gravitational lensing by large-scale structure generates from $E$ modes \citep{ZaldarriagaSeljak1998}. Part of this lensing $BB$ power can be removed with an estimate of the lensing potential, a procedure called delensing \citep{KnoxSong2002, Kesden2002, Delensing2021}. Today, the tightest limits on $r$ come from combining the fourth \textit{Planck} data release (PR4) with BICEP/Keck data \citep{BK18} and with baryon acoustic oscillation and CMB lensing data, which give $r < 0.032$ at $95\%$ confidence \citep{Tristram2022}. Moreover, how tightly $r$ can be bounded depends on the statistical framework and the priors, not only on the data \citep{Galloni2023, Galloni2024}. 

In the near future, SO SATs will reshape this knowledge. The nominal survey, three SATs observing for five years, forecasts an error $\sigr = (1.9$-$3.1)\times10^{-3}$ for $r = 0$ with half of the lensing $BB$ power removed \citep{SOsat2025}, where the range spans their noise and foreground assumptions. With six SATs observing until 2035, the forecast becomes $(0.9$-$1.4)\times10^{-3}$ at the same delensing, and $(0.7$-$1.2)\times10^{-3}$ if $70\%$ of the lensing power is removed \citep{SOsat2025}. Right behind that horizon, the LiteBIRD satellite mission will also join the search, targeting the same observable from space with full-sky coverage \citep{LiteBIRDinflation, LiteBIRDoverview}.

Many ground-based experiments do not analyse the sky they observe, but the one that survives their filtering. For example, atmospheric and instrumental $1/f$ noise force the removal of long time-scale modes from their time-ordered data (TOD). This filtering is a standard and necessary step in their pipelines, applied since the first degree-scale $B$-mode analyses \citep{BICEP1filtering, BICEP1threeyear, SPTpolBB}. When the filtering and the map-making are linear, their composition is the observation matrix (the linear operator that maps the true sky onto the filtered map) \citep{Pryke2009, TODfiltering, ObsMatrix}.

The standard treatment of a filtered map is to estimate power spectra with a \pcl method, which measures the power spectrum of the masked map and corrects it both for the multipole coupling that the mask induces \citep{MASTER, Alonso2019} and for the filtering, with a transfer function measured from simulations. Indeed, those filters suppress the signal at large scales that needs to be recovered. When dealing with CMB polarisation, the problem is also complicated by the spin nature of the field. In fact, on a cut sky, $E$ and $B$ polarisation modes get mixed, and the latter gets contaminated by leakages from the former. The \pcl method has been extended to handle the $E/B$ ambiguity at harmonic level using the so-called purification \citep{Smith2006, Grain2009}. However, the correction is applied to the measured spectra once the estimator has already run, so the estimator weights the data as if no filter had been applied. On the other hand, using the observation matrix to separate $E$ from $B$ in the maps before the spectra are computed is a very efficient alternative \citep{ObsMatrix}, and the same matrix now also enters map-level component separation \citep{BKXX}.

This paper is the second in a series describing \cosmoforge, a framework for optimal CMB power spectrum estimation and pixel-based likelihood analysis. Paper~I \citep{CosmoForgeI} introduced the quadratic maximum-likelihood (QML) estimator \citep{Tegmark1997, TegmarkOliveira2001} that it implements, and the pixel-space Gaussian likelihood that shares its covariance model. In such a context, the power spectra can be recovered with the smallest error bars allowed by the data, and the likelihood can be evaluated exactly on the cut sky maps. 
In this work, we add the capability to handle arbitrary linear pixel-space filters and measure the error bars it gives. \cosmoforge now takes the filter as an input, so that an arbitrary linear pixel-space filter $\Rmat$ enters the QML estimator exactly by conjugation, which replaces every covariance-like object $\mathbf{X}$ of the estimator with $\Rmat\mathbf{X}\Rmat\tp$ \citep{TTH1997}. Thus, the filtered response and the filtered covariance live in the same likelihood, and no correction step is needed.

This paper compares that capability on a SAT-like footprint representative of a ground-based experiment with two state-of-the-art \pcl pipelines: the pure \pcl estimator, with harmonic $B$-purification \citep{Smith2006, Grain2009} and a transfer function measured from simulations, and the matrix-based purification of BICEP/Keck \citep{ObsMatrix}, built from the same filter and fiducial spectra as QML, which we call matrix-purified \pcl. Since the performance of \pcl depends strongly on mask apodisation, i.e. tapering the mask weight smoothly to zero at its edge, on purification and weighting, both pipelines are tuned in every filter state and at every noise level (Sect.~\ref{sec:methods:protocol}).

We show results in different metrics, but the most relevant for the $r$ search is the ratio of error bars $\sigr$ between two estimators, which we also place against the pixel-likelihood bound, the smallest error bar any estimator can reach on the same maps (Sect.~\ref{sec:results}).

This work is organised as follows: Sect.~\ref{sec:methods} defines the patch, simulations, filters, QML, the two \pcl pipelines and the comparison protocol. Section~\ref{sec:results} presents the comparison on unfiltered maps and under a harmonic high-pass filter, an $\ell<30$ projector, an anisotropic scan-like filter, and a horizon-fixed ground template. The last two are built as constant-elevation scans either at the South Pole or at an Atacama-like site. We also measure how far each estimator sits from the pixel-likelihood bound. Section~\ref{sec:conclusions} states what is deliberately idealised alongside the conclusions. Five appendices carry the supporting material: the construction of the filters, with the toy scan strategy behind the constant-elevation ones (Appendix~\ref{app:filters}), the structural checks on the restricted estimator (Appendix~\ref{app:checks}), the unfiltered \pcl configuration search and the analytic covariances (Appendix~\ref{app:unfiltered}), the matrix transfer function as a bias correction (Appendix~\ref{app:tf}), and the pixel likelihood in closed form together with its controls (Appendix~\ref{app:pencil}).

\section{Method}
\label{sec:methods}

\subsection{Sky patch, simulations, instrument model}
\label{sec:methods:sims}

\begin{figure*}[t]
  \centering
  \includegraphics[width=\textwidth]{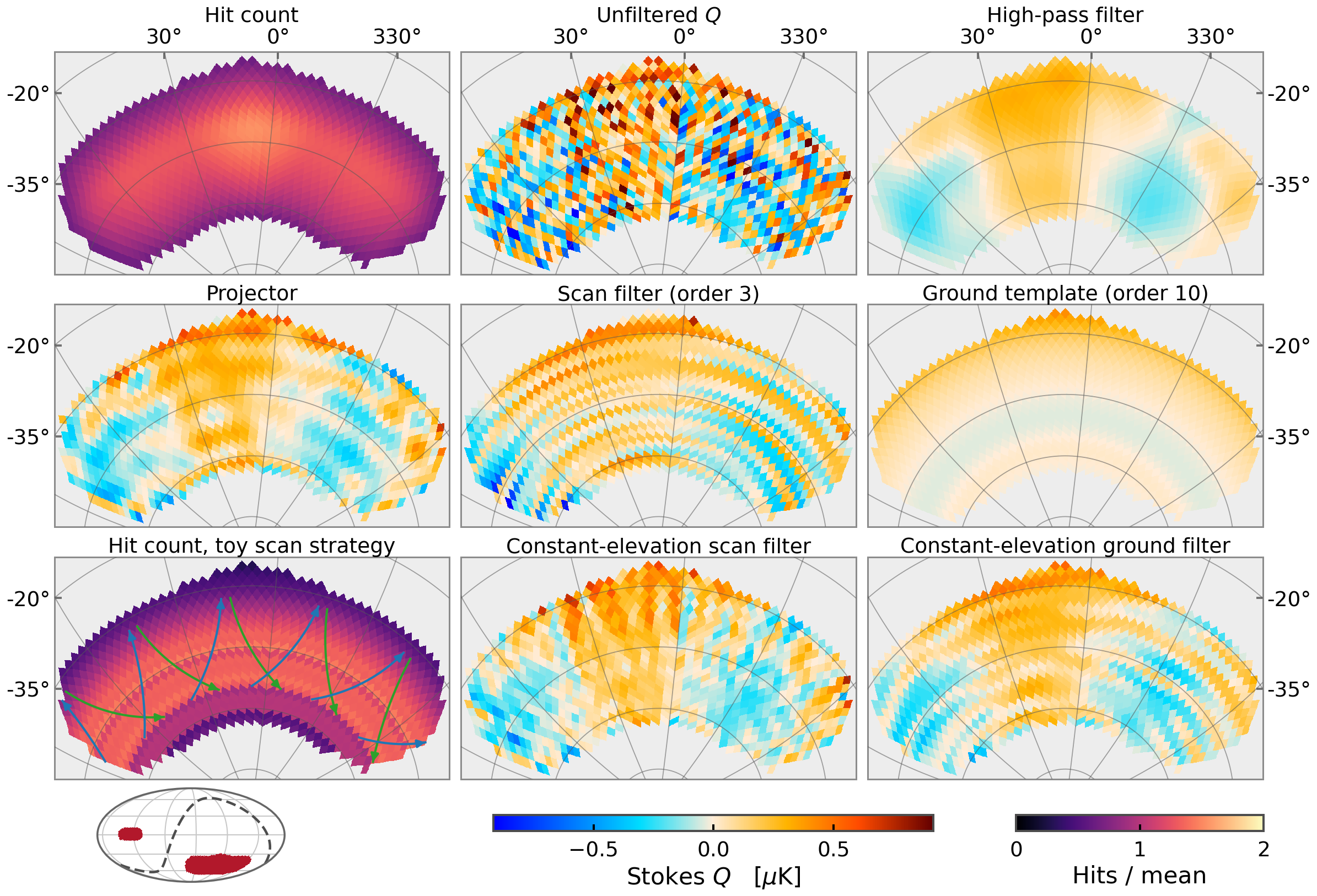}
  \caption{The southern field of the SAT patch and what the filters remove. The $\fsky = 0.10$ footprint is two disjoint fields, and the various panels show the southern one. The full-sky map at the bottom (equatorial coordinates, Galactic plane dashed) shows their relative position. These panels use a Lambert equal-area projection centred on the field with graticule lines every $30^\circ$ in right ascension and $15^\circ$ in declination; grey is outside the footprint. Top row: the SAT hit-count weight relative to its mean, one Stokes $Q$ realisation (signal plus noise), and the component that the harmonic high-pass filter removes from it. Middle row: the components removed by the projector, the scan filter at polynomial order 3, and the horizon-fixed ground template at order 10. Bottom row: the hit count of the toy constant-elevation scan strategy relative to its mean, and the components removed by the constant-elevation scan and ground filters at the same orders (Sect.~\ref{sec:methods:filters}). The arrows are the central detector sweeps illustrating the toy scanning strategy, with rising (green) and setting (blue) scans.}
  \label{fig:maps_filters}
\end{figure*}

As a representative of a ground based experiment, we use the publicly available SO SAT normalised hit map, the relative number of observations per pixel, released with \citet{SOforecast} and used by \citet{Wolz2024}\footnote{\url{https://portal.nersc.gov/cfs/sobs/users/so_bb/norm_nHits_SA_35FOV_ns512.fits}}, degraded to the working resolution and thresholded to a sky fraction $\fsky = 0.10$ (Fig.~\ref{fig:maps_filters}, top left). The patch is pixelised following the \texttt{HEALPix} scheme \citep{Gorski2005} at resolution parameter $N_{\rm side} = 32$, which gives $n_{\rm pix} = 1228$ pixels, that is $2n_{\rm pix} = 2456$ rows in the stacked Stokes $[Q\,|\,U]$ pixel vector. The footprint consists of two disjoint fields (Fig.~\ref{fig:maps_filters}), a southern one and an equatorial one holding $21\%$ of the pixels. All operators in this work, filters, covariances, and bandpower derivatives, are matrices on that vector.

In the philosophy of this work, the patch represents what we are handed by a generic experiment after some observation and a perfect component separation. Thus, QML and \pcl receive the same pixels and nothing else. QML uses all of them, while \pcl must apodise within them. Since the footprint edge sits where the hit count is still high, any taper removes pixels that carry weight, so the cost of apodisation measured in Sect.~\ref{sec:results:ratio} is specific to this footprint. Apodisation is then one of the \pcl configuration knobs rather than a property of the patch. QML's native analogue of that knob is the invertible diagonal weighting $\Rmat = \mathrm{diag}\!\left(\sqrt{\mathrm{hits}/\mathrm{median}}\right)$ (with the median taken over the footprint), which we show below leaves QML unchanged.

Signal realisations are drawn from \texttt{CAMB} lensed-scalar spectra at $r=0$ \citep{Lewis2000}, in a fiducial cosmology set to the \textit{Planck} 2018 TT,TE,EE+lowE+lensing marginalised means \citep{Planck2018VI}, $H_0 = 67.36\,\mathrm{km\,s^{-1}Mpc^{-1}}$, $\Omega_b h^2 = 0.02237$, $\Omega_c h^2 = 0.1200$, $\sum m_\nu = 0.06\,\mathrm{eV}$, $\Omega_k = 0$, $\tau = 0.0544$, $A_s = 2.1\times10^{-9}$, $n_s = 0.9649$, computed up to multipole $1500$. The tensor template used for $\sigr$ is a separate \texttt{CAMB} run at $r=1$ with tensors enabled and assumed to be scale-invariant (zero tensor spectral tilt), so that $r$ enters linearly, which simplifies the analysis. The template is the tensor $BB$ spectrum restricted to $30 \le \ell \le 64$, the window in which we infer $r$, and it is the same for every method of this work (Sect.~\ref{sec:methods:pics}). Sky realisations are convolved with the beam, truncated at the working band limit $\ell_{\rm sig} = 95 = 3N_{\rm side} - 1$ and pixelised with no pixel window, so that we can describe exactly the signal covariance (see  Eq.~\eqref{eq:sigma}). SAT analyses reach multipoles of a few hundred \citep{Wolz2024}; here, however, we work at $N_{\rm side} = 32$ to limit the computational cost, and Appendix~\ref{app:checks:binning} checks the effect of moving to $N_{\rm side} = 64$.

We assume a beam of $30'$ full-width at half-maximum, that of the SO SATs at 93\,GHz \citep{Wolz2024}, and most of this work assumes a white-noise map depth of $2\ukam$. We also run some tests at higher noise: $3.5$, $5$, and $10\ukam$, where the noise dominates the band. A last test adds an optimistic $1/f$ increment to the $2\ukam$ white level. The white noise at every depth is the same set of draws rescaled, so all depths share their realisations. These configurations exist to separate what is a property of the estimators from what is a property of how much signal the band carries (Sect.~\ref{sec:results:ratio}). For the $1/f$ case, we use the angular power spectrum $N_\ell = N_{\rm white}[1 + (\ell/\ell_{\rm knee})^{\alpha}]$ to describe the noise, with $\ell_{\rm knee} = 25$ and $\alpha = -2.5$, the optimistic 93\,GHz values of \citet{Wolz2024}. Noise is added in pixel space, uncorrelated between pixels, with variance $\sigma_p^2 \propto 1/\mathrm{hits}_p$ normalised so that the quoted map depth holds at the mean hit count.

Every simulation set used in this analysis is composed of 1000 realisations. Aside from the simulations described just above, we produce some extra sets used for specific parts of the analysis. For example, we need noiseless $E$-mode, lensing $B$-mode, and tensor $B$-mode simulations to get the expectation values of bandpowers under different filters and different estimators. We draw three independent realisation sets per case, each with its own signal and noise draws, and the scatter between them gives the Monte Carlo uncertainty of each comparison (Sect.~\ref{sec:methods:protocol}).

\subsection{Filters}
\label{sec:methods:filters}

Every filter in this work is a linear operator $\Rmat$ on the $QU$ pixel rows of the footprint (Sect.~\ref{sec:methods:sims}). Here, we build a ladder of filters to showcase their effect on the power spectrum estimation and $r$ inference. They are not meant to be realistic, but instead they represent different classes of filters. Figure~\ref{fig:maps_filters} shows what each of them removes from a $Q$ map, and Appendix~\ref{app:filters} gives their detailed construction.

Firstly, we have two isotropic filters acting on $E$ and $B$ identically. One is a harmonic high-pass filter that suppresses both smoothly below $\ell_0 = 30$ (Eq.~\eqref{eq:softfilter}). It is invertible, so it rescales every mode rather than deleting any. The other is a projector that nulls the harmonics for $\ell < 30$ and is the aggressive counterpart of the high-pass.

Then, we define two anisotropic filters that mimic what a ground-based telescope does to its time streams. For instance, it is quite typical of ground telescopes to remove a polynomial of low order along each scan. BICEP1 and BICEP2 remove a third-order polynomial \citep{BICEP1threeyear, BICEP2instrument}; SPTpol on the South Pole Telescope \citep{SPTpolBB} uses a fifth- to ninth-order polynomial, while the filtered SAT simulations of \citet{SOtransfer} fit one of third order in time to each subscan. Our scan filter removes the same polynomials, at orders $n = 1, 3, 5$, along each stripe of constant declination. On the other hand, signal fixed in horizon coordinates, such as ground emission picked up through the far sidelobes of the beam, is instead removed by a template in azimuth, which in map space removes sky structure constant in right ascension. We name this ground filter, and we use it at orders $n = 5, 10, 15, 25$. The scan filter is dense and removes hundreds of modes, while the ground filter removes only tens.

Both filters remove whole rings, or stripes, of constant declination, which is the geometry a constant-elevation scan traces at the South Pole. This is helpful to isolate the effect of filtering along the two directions separately. On the contrary, it is also interesting to check what happens for a more complex geometry. Indeed, from an Atacama-like site \citep{SOtransfer}, a scan at fixed elevation cuts across parallels of declination, and rising and setting scans cross each other on the sky. We therefore build both filters again, at orders 3 and 10, by filtering and binning the time streams of a toy constant-elevation scan strategy. The ground filter is then a Legendre polynomial of order 10 in azimuth fitted to each scan, similar to the ground filter of the filtered SAT simulations of \citet{SOtransfer}, which they fit per observation together with the time-domain polynomial. Since a pattern removed by one pass is only partly removed by the others, these filters attenuate most modes rather than removing them (see Appendix~\ref{app:filters} for more details).

Together with the unfiltered maps, the eleven filters of the ladder give a total of twelve states. The stripe filters act on $Q$ and $U$ independently in the equatorial frame. With a rotating half-wave plate, a real filter acts in the instrument frame, which rotates against the sky along a scan, and therefore also mixes $Q$ and $U$. We neglect the rotation for the stripe filters to keep them as simple as possible. On the other hand, the constant-elevation filters include this effect.

\subsection{QML and the exact injection of a filter}
\label{sec:methods:qml}

We use the \qube implementation of the QML estimator from the \cosmoforge framework (Paper~I)\footnote{\url{https://github.com/ggalloni/CosmoForge}}; we recall here only what the injection of a filter needs. The following arguments hold for any choice of fields and bandpowers, but here we specify them for the analysis of this work. Write the data as a single vector $\dvec = \svec + \nvec$ over the $2 n_{\rm pix}$ stacked $[Q\,|\,U]$ pixels of the patch, with covariance $\Cmat = \Smat(\Cl) + \Nmat$. Here $\Nmat$ is the known noise covariance, either diagonal or non-diagonal ($1/f$ case), and $\Smat$ is linear in the bandpowers being estimated:
\begin{equation}
  \Smat(\Cl) \;=\; \sum_\ell \Cl\, \Pmat_\ell.
  \label{eq:sigma}
\end{equation}
Here, $\Pmat_\ell = \partial \Cmat/\partial \Cl$ is the derivative of the signal covariance with respect to the power spectrum at multipole $\ell$. As in Paper~I, it is built from the Legendre polynomials of the pixel-pair separations and carries the beam window $B_\ell^2$, since the simulated maps carry no pixel window (Sect.~\ref{sec:methods:sims}). The sum defining $\Smat$ runs to the imposed band limit of the modelled signal.

The estimator \citep{Tegmark1997,TegmarkOliveira2001,BondJaffeKnox1998} forms, for each bandpower $b$, the power in one bin of multipoles,
\begin{equation}
  y_b \;=\; \tfrac{1}{2}\left[\,
    \dvec\tp \Cmat^{-1} \Pmat_b \Cmat^{-1} \dvec
    \;-\; \mathrm{tr}\!\left(\Cmat^{-1} \Nmat \Cmat^{-1} \Pmat_b\right)\right],
  \label{eq:qml_y}
\end{equation}
with, in the notation of Paper~I, the bandpower derivative $\Pmat_b = \partial \Cmat/\partial C_b = \sum_{\ell \in b} \Pmat_\ell$ with respect to the bandpower $C_b$ of bin $b$, and the second term being the noise bias. The noise bias is a constant offset of the bandpowers and cannot change $\sigr$, so in this work we drop it, for QML as for \pcl. Finally, the Fisher matrix, which converts $y_b$ into the physical bandpowers we are familiar with, reads
\begin{equation}
  \Fmat_{bb'} \;=\; \tfrac{1}{2}\,
  \mathrm{tr}\!\left(\Cmat^{-1} \Pmat_b \Cmat^{-1} \Pmat_{b'}\right),
  \qquad
  \hat C_b \;=\; \sum_{b'} \left(\Fmat^{-1}\right)_{bb'} y_{b'}.
  \label{eq:qml_fisher}
\end{equation}
Thus, $\hat C_b$ is unbiased with covariance $\mathrm{Cov}(\hat C_b, \hat C_{b'}) = (\Fmat^{-1})_{bb'}$, which ensures the estimator is optimal and is available analytically with no simulations\footnote{Paper~I showed that $(\Fmat^{-1})_{bb'}$ matches the empirical covariance of 1000 simulations at every multipole and sky fraction they tested. Table~\ref{tab:cov_fidelity} shows the same on this patch.}. Binning is native, $\Pmat_b$ replacing $\Pmat_\ell$ inside Eqs.~\eqref{eq:qml_y}-\eqref{eq:qml_fisher}, so the deconvolved bandpower $\hat C_b$ is a Fisher-weighted combination of the $\Cl$ within the bin rather than a flat average (Paper~I). Note that the estimator is dense: one solve costs $\mathcal{O}(n_{\rm pix}^3)$ for the $2n_{\rm pix} = 2456$ map rows here. That sets the resolution this comparison runs at ($N_{\rm side} = 32$), and it is why \pcl, cheaper than QML, can be re-searched in every state and at every depth.

An arbitrary linear filter $\Rmat$ acting on the map vector is injected by conjugation of every object in the estimator, with no change to the estimator internals:
\begin{equation}
  \dvec \to \Rmat \dvec, \quad
  \Cmat \to \Rmat \Cmat \Rmat\tp, \quad
  \Nmat \to \Rmat \Nmat \Rmat\tp, \quad
  \Pmat_b \to \Rmat \Pmat_b \Rmat\tp.
  \label{eq:conjugation}
\end{equation}
Because the filtered bandpower derivatives and the filtered covariance enter the same likelihood, Eqs.~\eqref{eq:qml_y}-\eqref{eq:qml_fisher} go through unchanged, and the estimator remains unbiased wherever the filter leaves information, with no transfer-function step. For invertible $\Rmat$, nothing is lost at all: the Fisher information of a Gaussian is invariant under any invertible linear map of the data, which enters the traces of Eq.~\eqref{eq:qml_fisher} as a similarity transform and cancels \citep{TTH1997}; the same cancellation leaves $y_b$ unchanged. The conjugated path is therefore algebraically identical to the unfiltered one, and the invertible hits-weighting operator applied to this case reproduces it. This is an internal test our implementation passes.

For singular $\Rmat$, the conjugated covariance is singular too, as \citet{ObsMatrix} note for the observation matrix of BICEP/Keck. We then have two options: either regularise the covariance back to full rank, or run the estimator in the range of the filter. The first is a ridge-regularised conjugation, which adds a finite variance, the ridge, to the directions the filter removes. The second restricts the analysis to the directions the filter keeps, through a singular value decomposition (SVD) of $\Rmat$. The two agree, since the directions carrying the ridge are the ones the restriction drops, as for the marginalisation over templates by infinite variance \citep{BondJaffeKnox1998, Slosar2004}, and Appendix~\ref{app:checks:restriction} verifies this over six orders of magnitude in the ridge amplitude. \qube implements the second, which needs no regularisation, and every conjugated QML number in this work is computed with it.

Specifically, we keep the $n_{\rm R}$ singular values of $\Rmat$ above a fraction $\epsilon$ of the leading one, and write the truncated SVD $\Rmat \simeq \Umat\Dmat\Zmat\tp$. Here, $\Dmat$ is the $n_{\rm R} \times n_{\rm R}$ diagonal matrix of those singular values, while $\Umat$ and $\Zmat$ are $2n_{\rm pix} \times n_{\rm R}$ and have orthonormal columns, which span the directions the filter outputs and the directions of its input that it keeps, respectively. The analysis then runs in those $n_{\rm R}$ dimensions: the filtered data vector becomes $\Umat\tp\Rmat\dvec$, of length $n_{\rm R}$, and every filtered covariance-shaped object becomes the $n_{\rm R} \times n_{\rm R}$ matrix $\Umat\tp(\cdot)\Umat$. Since $\Dmat$ is an invertible diagonal matrix, the estimator is unchanged by its presence, and nothing is divided by a small singular value. All the filters of this work are symmetric with non-negative eigenvalues, so that $\Umat = \Zmat$ and the SVD reduces to their eigendecomposition. A projector is the special case where, in addition, $\Dmat = \Imat$, and this becomes a plain subspace restriction. Thus, the rank threshold $\epsilon$ is the one declared quantity that controls which modes the analysis treats as removed; hence, it is a parameter of \qube. We use $\epsilon = 10^{-2}$ in every filter state.

Note that, beyond the noise bias, Eq.~\eqref{eq:qml_y} subtracts nothing, so signal outside the estimated range $[\ell_{\min}, \ell_{\max}]$ leaks into every bandpower as the additive term $\tfrac{1}{2}\mathrm{tr}(\Cmat^{-1} \Smat_{\rm out} \Cmat^{-1} \Pmat_b)$, with $\Smat_{\rm out}$ the signal covariance of the multipoles outside that range. That term left unaccounted for is large enough to be mistaken for an estimator bias (Paper~I). Here, to deal with that, we use a buffered band definition, where the estimated range is $[\ell_{\min}, \ell_{\max}]$ but the reference theory is evaluated over a wider range to absorb the extra signal. Buffered bandpowers are estimated jointly with the reported ones, in bins of $\Delta\ell = 5$ over $20 \le \ell \le 29$ and $65 \le \ell \le 94$ around the reported $30 \le \ell \le 64$, and then discarded, so that the out-of-band term of Eq.~\eqref{eq:qml_y} has somewhere to go. On the high side, the signal survives every filter, so a buffer is always needed. On the low side, the need depends on the filter (Appendix~\ref{app:checks:buffer}), but we keep the buffer in every state so that all states are compared over one band.

Appendix~\ref{app:checks} reports the invertibility and ridge tests above, and two further checks: that the nulled subspace is affordable, and that a bin the filter destroys shows a large variance rather than a quiet bias. It also measures the cost of conjugating only part of the QML algebra, and the effect of changing $\Delta\ell$ or the resolution to $N_{\rm side} = 64$.

\subsection{Two \pcl pipelines}
\label{sec:methods:pcl}

\subsubsection{Harmonic and matrix $E/B$ separation}

Pseudo-$C_\ell$ spectra are computed with \texttt{NaMaster} 2.6 \citep{Alonso2019}\footnote{\url{https://github.com/LSSTDESC/NaMaster}. No significant difference is found when rerunning with version 3.0.}. The masked-sky (pseudo) spectra of the weighted map are related to the true ones by a mode-coupling matrix determined by the mask and the weighting \citep{MASTER},
\begin{equation}
  \begin{aligned}
  \bigl\langle \tilde C_\ell^{\,X} \bigr\rangle
  &\;=\; \sum_{\ell'} \sum_{Y}
  M^{XY}_{\ell\ell'} \, B^2_{\ell'} \, C^{\,Y}_{\ell'}
  \;+\; \tilde N^{\,X}_\ell ,\\
  \hat C^{\,X}_b &\;=\; \sum_{b'} \sum_{Y}
  \left(\mathcal{M}^{-1}\right)^{XY}_{bb'}
  \left(\tilde C^{\,Y}_{b'} - \tilde N^{\,Y}_{b'}\right),
  \end{aligned}
  \label{eq:master}
\end{equation}
where $\tilde C^X_\ell$ and $\tilde N^X_\ell$ are the pseudo-spectra of the map and of the noise, $X, Y$ run over the polarisation spectra and $\mathcal{M}$ is the binned coupling matrix. In polarisation, $\mathcal{M}$ is not block diagonal in $X$: a cut sky mixes $E$ and $B$ through the ambiguous modes. \texttt{NaMaster}'s $B$-purification removes the mask-induced part of that mixing, at the cost of differentiating the mask, which is why it works only jointly with a resolved apodisation taper (Appendix~\ref{app:unfiltered:mask}). However, a filter can also mix $E$ into $B$, and such a mixing is already in the map \texttt{NaMaster} receives, so purification cannot remove it; hence, a \pcl analysis can correct it only in the mean, through the transfer function introduced below \citep{SOtransfer}. Removing it at the map level requires the observation matrix itself, as in the matrix-based purification of \citet{ObsMatrix}, who build that matrix by mirroring, step by step, the map-making pipeline developed for the QUaD experiment \citep{Pryke2009}. \citet{SOtransfer} make the same point and describe building that matrix at the resolution of SAT maps as an area of active research.

We therefore run \pcl in two forms, which differ in how $E$ and $B$ are separated. The first is the pure \pcl estimator \citep{Smith2006, Grain2009}, as implemented by \texttt{NaMaster} with $B$-purification, applied to the filtered maps and followed by a transfer function measured from filtered simulations, which we take in the matrix form below; we call this pipeline pure \pcl. The SO SAT analyses have validated the two components separately: purification on unfiltered maps \citep{Wolz2024} and the transfer function on filtered maps without purification \citep{SOtransfer}. The second separates $E$ from $B$ in pixel space before \texttt{NaMaster}, with the filter known, following the matrix-based purification of BICEP/Keck \citep{ObsMatrix}, which extends the pixel-space eigenmode method of \citet{Bunn2003}. \citet{Lewis2002} give the corresponding construction in harmonic space. Indeed, writing $\Smat_E$ and $\Smat_B$ for the signal covariances of Eq.~\eqref{eq:sigma} built from the fiducial lensed $EE$ and lensing $BB$ spectra alone, it solves the generalised eigenproblem
\begin{equation}
  \Rmat \Smat_B \Rmat\tp \mathbf{q} \;=\; \lambda \left(\Rmat \Smat_E \Rmat\tp + \eta \Imat\right) \mathbf{q} ,
  \label{eq:purifier}
\end{equation}
and keeps the eigenvectors $\mathbf{q}$ with generalised eigenvalue $\lambda$ above a threshold $\lambda_{\rm c}$. Here, $\eta$ is a ridge on the $E$ covariance, set to $10^{-4}$ of the mean diagonal of $\Rmat \Smat_E \Rmat\tp$, which regularises the directions where that covariance vanishes. With this construction, every direction in the span of the kept eigenvectors carries at least $\lambda_{\rm c}$ times more $B$ than $E$ power. This is similar to the construction of \citet{ObsMatrix}\footnote{\citet{ObsMatrix} use $C_\ell \propto \ell^{-2}$ for both covariances, regularise both sides with $1\%$ of the mean diagonal, and keep the quarter of the directions with the largest eigenvalues. In our construction, since the kept directions carry almost no $E$ power, $\lambda_{\rm c}\eta$ also acts as a floor on the $B$ power a kept direction must carry, like a noise term in the $E$ covariance.}. With $\mathbf{G}$ the $2n_{\rm pix} \times n_{\rm G}$ matrix whose orthonormal columns span the $n_{\rm G}$ kept eigenvectors, the purified map is $\mathbf{G}\mathbf{G}\tp\Rmat\dvec$, which \texttt{NaMaster} then analyses with its own purification switched off. The purifier is a dense generalised eigenproblem on the filtered pixel vector and needs the filter as a matrix, as QML does. We use $\lambda_{\rm c} = 10$, which keeps $n_{\rm G} = 584$ to $777$ of the $2456$ directions depending on the filter, and Appendix~\ref{app:checks:thresholds} varies it. The fiducial spectra, the beam, and $\Rmat$ are the ones QML is built from, so the two methods receive the same information. Note that we use \cosmoforge to obtain the signal covariances necessary for this pipeline, which we call matrix-purified. Thus, not only does \cosmoforge provide the QML and pixel-based implementations, but it also delivers the inputs for the matrix-purified \pcl. 

\subsubsection{Configuration search}

As mentioned above, the \pcl estimation needs to be tuned to provide its best estimate. The configuration space we search comprises the apodisation scale, the apodisation kernel, the purification choice, and the pixel weighting:
\begin{itemize}
\item apodisation scale, the width of the taper: a binary mask plus $\{3, 5, 7.5, 10, 15, 20\}$ degrees;
\item kernel: the two taper profiles of \texttt{NaMaster}, \texttt{C1} and \texttt{C2};
\item purification: none, $B$ only, or $E$ and $B$;
\item weighting: uniform, or inverse-noise (the hit map).
\end{itemize}
Note that the axes are not independent: for example, purification demands some apodisation. Still, the grid keeps the unapodised purified configurations rather than excluding them so that they can measure that coupling (Table~\ref{tab:sat_sweep}, lower block). In total, this gives 78 configurations. Since the $\sigr$ study is a $BB$-only measurement, there the pure \pcl grid is reduced to 20: purification fixed to $B$ only, since purifying $E$ cannot change a $BB$ number, the unpurified configurations dropped as a factor several behind (Table~\ref{tab:sat_sweep}), and the binary and $3^\circ$ masks dropped as unable to support purification, leaving tapers $\{5, 7.5, 10, 15, 20\}$ degrees, both kernels and both weightings. Since the matrix-purified maps retain only about $10^{-3}$ as much $E$ as $B$ power, they need no harmonic purification, and their grid reaches down to the binary mask. It comprises the binary mask with both weightings and the $\{3, 5, 7.5, 10\}$ degree tapers with both kernels and both weightings, 18 configurations in all.

\subsubsection{Transfer functions}

For filtered maps, \pcl receives the standard Monte Carlo treatment \citep{MASTER, SOtransfer}. So, we measure a transfer function from signal-only filtered simulations and apply it as a full $2 n_b \times 2 n_b$ matrix over the stacked $[EE\,|\,BB]$ bandpowers of the $n_b = 15$ bins over $20 \le \ell \le 94$,
\begin{equation}
  \bigl\langle \hat{C}^{X,\rm filt}_b \bigr\rangle
  \;=\; \sum_{Y} \sum_{b'} T^{XY}_{bb'}\, C^{Y,\rm true}_{b'} ,
  \qquad X, Y \in \{EE, BB\},
  \label{eq:tf}
\end{equation}
so that $E \to B$ leakage is corrected in the mean through the off-diagonal block $T^{BB\,EE}_{bb'}$ of $\Tmat$. Each column of $\Tmat$ is measured from a delta-bin probe carrying unit power in a single bin of a single spectrum, $2 n_b = 30$ columns in all, with $1000$ signal-only realisations each for a total of 30,000 simulations. As in \citet{SOtransfer}, we measure them on coupled binned pseudo-spectra rather than on decoupled bandpowers, which are ill-conditioned on a patch this size. Appendix~\ref{app:tf:published} compares this construction with published ones \citep{HWPmodulation, SOtransfer}.

The transfer function is a linear correction to the bandpowers, invertible on the full $[EE\,|\,BB]$ vector, so the Fisher invariance argument of Sect.~\ref{sec:methods:qml} applies \citep{TTH1997}, and it cannot add information on $r$. The $\sigr$ of this work uses only the reported $BB$ bins, while $\Tmat^{-1}$ mixes $EE$ and buffered bins into them, so we also measure its effect there (Appendix~\ref{app:tf}).

The matrix-purified bandpowers are also biased low, since the purification removes part of the $B$ response. Thus, this pipeline also needs a transfer function to give unbiased bandpowers, measured from simulations or computed from the observation matrix as in \citet{ObsMatrix}. Since its $E \to B$ leak is small (Sect.~\ref{sec:results:leak}), this transfer function needs no $BB \leftarrow EE$ block, but it still has to be built for every filter and configuration. We measure it for the per-bandpower comparison of Sect.~\ref{sec:results:bandpowers}. Its $\sigr$ is unaffected, since each estimator is scored against its own measured response (Sect.~\ref{sec:methods:protocol}).

\subsection{The pixel likelihood as an optimality anchor}
\label{sec:methods:pics}

We add a third method to this comparison, which performs no data compression at all. This allows us to place the bandpower estimators against an absolute scale rather than only against each other: the pixel-space Gaussian likelihood that \pics implements in \cosmoforge (Paper~I), evaluated in closed form (Appendix~\ref{app:pencil}) from the signal matrices of \qube and on the identical simulations, noise model, mask, beam, and filter operators,
\begin{equation}
  \begin{aligned}
  -2\ln\mathcal{L}(r) &= (\Rmat\dvec)\tp \Cmat_{\Rmat}(r)^{-1} \Rmat\dvec + \ln\left|\Cmat_{\Rmat}(r)\right|,\\
  \Cmat_{\Rmat}(r) &= \Rmat\left(\Smat_{\rm s} + r\,\Smat_{\rm t} + \Nmat\right)\Rmat\tp.
  \end{aligned}
  \label{eq:pics_like}
\end{equation}
Here, $\Smat_{\rm s}$ is the $r=0$-scalar sky contribution to the signal covariance matrix ($EE$ plus lensing $BB$), while $\Smat_{\rm t}$ is the $r=1$ tensor $BB$ one. The curvature of Eq.~\eqref{eq:pics_like} at $r=0$ is the information the data carry, and it is analytic (see Appendix~\ref{app:pencil}). The gap between that and $\sigr(\mathrm{QML})$ is the cost of compressing the maps into the reported bandpowers, which includes their marginalisation over the buffered and $EE$ bandpowers, and the gap to $\sigr(\text{\pcl})$ sums on top of that the cost of apodisation, weighting, and harmonic $E/B$ separation. Replacing $\Rmat$ by $\mathbf{G}\tp\Rmat$ in Eq.~\eqref{eq:pics_like} gives the same bound for the matrix-purified data, that is, the information purification leaves to any estimator.

We emphasise that the likelihood has no bandpowers, but it has an inference window in harmonic space. Multipoles outside $30 \le \ell \le 64$ are held at the fiducial $r = 0$ spectrum, so that the inference is limited there while we still represent the covariance of the map fully. Also, since the same algebra serves both \qube and \pics, the same filtered covariance is used in both, ensuring full coherence between the two.

\subsection{Comparison and figures of merit}
\label{sec:methods:protocol}

To ensure the fairness of any comparison, expectation values of bandpowers from each estimator are measured from the signal-only sets of Sect.~\ref{sec:methods:sims}, and each estimator is scored against its own. In addition, as mentioned above, each \pcl grid is searched identically per filter state, per depth, and per realisation set, and we always quote it at its optimum (except when explicitly stated). Indeed, without this search, the tuning would be a dominant systematic of the comparison. For example, pinning the unfiltered state to the pure \pcl configuration that is optimal under the order-3 scan filter gives a $\sigr$ ratio of $1.45$, where the state's own optimum gives $1.21$. Once \pcl is tuned per state, the main source of uncertainty in this work is Monte Carlo noise. Within a realisation set, every method sees the same skies, so all comparisons are matched realisation by realisation. This is why every number quoted here is an average over three sets of realisations (see Sect.~\ref{sec:methods:sims}), and the scatter between sets is at most $3\%$ on the pure \pcl ratios and $1.2\%$ on the matrix-purified ones along the ladder. Where a table, a column, or a number comes from a single set, the caption or the text says so.

As regards figures of merit, our main one is $\sigr$, the error on the amplitude of the tensor template of Sect.~\ref{sec:methods:sims}, with the amplitude of the lensing $BB$ spectrum held fixed. Let $\mathbf{t}_0$ be the measured mean $BB$ response to the tensor template and $\boldsymbol{\Sigma}$ the empirical $BB$ bandpower covariance over $n_{\rm sim} = 1000$ realisations. Then, from the Fisher matrix of a Gaussian likelihood with a parameter-dependent mean and a fixed covariance \citep{TTH1997},
\begin{equation}
  \sigma(r)^{-2} \;=\; \mathbf{t}_0^{\mathsf T}\,\boldsymbol{\Sigma}^{-1}\,\mathbf{t}_0 .
  \label{eq:sigmaA}
\end{equation}
When we need to quote an absolute number, we correct the finite-sample bias of $\boldsymbol{\Sigma}^{-1}$ \citep{Hartlap2007} by rescaling it by $(n_{\rm sim} - n_{\rm fit} - 2)/(n_{\rm sim} - 1)$, with $n_{\rm fit}$ the number of bandpowers in the fit. In a ratio of error bars, both would be affected by the same rescaling factor, which cancels out. Then, this matters only where an empirical error is compared with an analytic one, against the pixel-likelihood bound in Sect.~\ref{sec:results:bound}.

We also compare the estimators bandpower by bandpower. Following Paper~I, we quote for each reported $BB$ bin the error on its amplitude in two forms. The marginal error is the square root of the diagonal of $\boldsymbol{\Sigma}$. The conditional error, $[(\boldsymbol{\Sigma}^{-1})_{bb}]^{-1/2}$, is the error on that bandpower when the others are held at their measured values. Since the two forms differ only through the off-diagonal terms of $\boldsymbol{\Sigma}$, comparing them measures how correlated the bandpowers are. In addition, both are divided by the estimator's own mean response to the lensing sky in that bin, so that we can compare them between estimators. 

Alongside these figures of merit, we report two further quantities a real analysis must live with. The first is the measured $E \to B$ leakage, i.e. the $BB$ bandpower an $E$-only sky produces in units of the $BB$ response to the lensing sky. To collapse that into a number, we average the absolute leakage power over the inference window $30 \le \ell \le 64$, and then divide by the average lensing power over the same range. Since this statistic sums absolute values, it is positive even for an estimator with no leakage at all. We therefore quote with it its Monte Carlo floor, the value it takes in expectation for zero leakage given the scatter of the per-bin means across the three realisation sets. Furthermore, since a mean leak is a bias that simulations can correct, we express it as the shift it induces in the estimate of $r$ at fixed lensing amplitude, in units of $\sigr$, through the linear bias of the template fit \citep{HutererTakada2005, AmaraRefregier2008}, $\Delta r/\sigr = \mathbf{t}_0\tp\boldsymbol{\Sigma}^{-1}\bar{\mathbf{e}}\,/\,(\mathbf{t}_0\tp\boldsymbol{\Sigma}^{-1}\mathbf{t}_0)^{1/2}$, with $\bar{\mathbf{e}}$ the mean $BB$ bandpowers of the $E$-only sky and $\mathbf{t}_0$ the tensor response defined above. The second is the fidelity of each estimator's analytic covariance against its own empirical one, i.e., QML's $\Fmat^{-1}$ and \texttt{NaMaster}'s Gaussian bandpower covariance. When an analysis cannot afford to get that from simulations, those are the quantities it must trust (Appendix~\ref{app:unfiltered:cov}).

\section{Results}
\label{sec:results}

We now compare QML with the two \pcl pipelines on unfiltered maps and under each filter of Sect.~\ref{sec:methods:filters}. The unfiltered maps are the most favourable case for \pcl, and they set the baseline. In fact, without a filter, the gap between the estimators can only come from how each of them handles the footprint. Firstly, on a cut sky, the mask couples different multipoles and mixes $E$ with $B$. A \pcl analysis must choose how to taper and weight the pixels to limit both effects, while QML natively accounts for them through its covariance. The \pcl configuration search and the analytic covariances, which involve no filter, are collected in Appendix~\ref{app:unfiltered}.

\subsection{$\sigr$ with and without filters}
\label{sec:results:ratio}

\begin{table*}[t]
  \centering \small \setlength{\tabcolsep}{4pt}
  \caption{$\sigr$ along the filter ladder.}
  \label{tab:sigr_filtered}
  \begin{tabular}{lcccccccc}
    \toprule
    & & & & & \multicolumn{2}{c}{$2\ukam$} & \multicolumn{2}{c}{$3.5\ukam$} \\
    \cmidrule(lr){6-7}\cmidrule(lr){8-9}
    State & Modes & Bound $\times10^{3}$ & Destroys & QML/bound & Pure/QML & Matrix-purified/QML & Pure/QML & Matrix-purified/QML \\
    \midrule
    Unfiltered  &   0 & 1.896 & 1.00 & 1.02 & 1.21 & 1.01 & 1.16 & 1.01 \\
    \midrule
    High-pass   &   2\tablefootmark{a} & 1.897 & 1.00 & 1.02 & 1.23 & 1.02 & 1.17 & 1.02 \\
    Projector   & 514 & 2.684 & 1.42 & 1.04 & 1.56 & 1.03 & 1.46 & 1.03 \\
    Scan 1      & 172 & 1.978 & 1.04 & 1.02 & 1.89 & 1.01 & 1.64 & 1.01 \\
    Scan 3      & 336 & 2.070 & 1.09 & 1.02 & 2.46 & 1.01 & 2.09 & 1.02 \\
    Scan 5      & 504 & 2.178 & 1.15 & 1.02 & 3.04 & 1.02 & 2.50 & 1.02 \\
    Const. scan 3 &  22\tablefootmark{a} & 2.218 & 1.16 & 1.03 & 3.65 & 1.10 & 2.78 & 1.04 \\
    \midrule
    Ground 5    &  12 & 1.898 & 1.00 & 1.03 & 1.20 & 1.01 & 1.15 & 1.01 \\
    Ground 10   &  22 & 1.903 & 1.00 & 1.03 & 1.20 & 1.00 & 1.16 & 1.01 \\
    Ground 15   &  32 & 1.912 & 1.01 & 1.02 & 1.25 & 1.01 & 1.19 & 1.01 \\
    Ground 25   &  50 & 1.931 & 1.02 & 1.02 & 1.40 & 1.01 & 1.29 & 1.01 \\
    Const. ground 10 &   6\tablefootmark{a} & 2.033 & 1.07 & 1.03 & 1.58 & 1.04 & 1.39 & 1.05 \\
    \bottomrule
  \end{tabular}
  \tablefoot{Values are from $BB$ over $30 \le \ell \le 64$ at $N_{\rm side} = 32$, with the tensor template confined to the same window and the lensing amplitude fixed. ``Modes'' is the number of pixel-space directions the filter removes; ``Bound'' is the pixel-likelihood $\sigr$ (Sect.~\ref{sec:methods:pics}), and ``Destroys'' is the bound relative to its unfiltered value, the increase no estimator can avoid. ``QML/bound'' is QML's $\sigr$ relative to the bound in the same state, measuring the information loss in passing from pixels to bandpowers. The last four columns give $\sigr$ of each \pcl pipeline relative to QML's, which instead measures the loss of passing to a suboptimal estimator. The first row is the unfiltered state and acts as the baseline each number should be compared to. Here, ``Const.'' stands for constant-elevation; thus, it refers to the toy scanning strategy and filtering at an Atacama-like site. Note that we do not report the unfiltered results of that configuration as they are almost identical to the first row (each number is matched well within Monte Carlo scatter). Here, only ``Bound'' and ``Destroys'' carry no Monte Carlo scatter as the pixel-based $\sigr$ is analytic.\\
  \tablefoottext{a}{These filters remove no mode exactly; the entry counts the directions below the rank threshold $\epsilon = 10^{-2}$ of Sect.~\ref{sec:methods:qml}, which the analysis drops.}}
\end{table*}

\subsubsection{Along the filter ladder}

Table~\ref{tab:sigr_filtered} collects the $\sigr$ comparison for every filter state, including the unfiltered maps in its first row. Along each column, we can measure the effect of the data compression, from pixels to bandpowers, and the effect of a suboptimal estimator, from QML to the two incarnations of \pcl. Instead, each row can also be read against the first to get the effect of the filters. Let us focus first on the columns assuming $2\ukam$ (the rest is discussed in Sect.~\ref{sec:dep_on_noise}).

Without filtering, the $\sigr$ of QML is only $2\%$ above the pixel-likelihood bound, while that of pure \pcl is $21\%$ above QML's. With a finer pixelisation, the pure \pcl $\sigr$ falls by up to about $10\%$ (Appendix~\ref{app:checks:binning}). Most of the gain, and about half of it under the projector, is already there at a fixed taper, from a better-resolved harmonic purification, and the rest comes from the narrower taper that the finer pixels make affordable. On the other hand, matrix-purified \pcl is very much compatible with QML.

These ratios grow along the ladder: the isotropic filters raise it moderately, the projector more than the high-pass filter, since both act mostly at the bottom of the band. The anisotropic scan filter costs the pure \pcl pipeline the most, and more at every polynomial order. The matrix-purified pipeline instead stays within a few per cent of QML in the unfiltered state and under every stripe filter. Leaving the constant declination scans and focusing on the constant-elevation in an Atacama-like site, the gap grows even more due to the more complex geometry of the problem. Indeed, the toy scanning strategy we used tends to attenuate modes rather than destroying them (see ``Modes'' entries with respect to the stripe filters of the same order). In such a scenario, QML still sits within a few per cent of the bound, while the pure \pcl pipeline pays more than under the stripe filter of the same order. Also, the matrix-purified pipeline no longer matches QML, losing from a few to ten per cent above it.

The pure \pcl cost is not simply a matter of how many modes a filter removes. Indeed, already at order 1 the scan ratio exceeds that of the projector, although the scan filter then removes a third of the modes the projector removes and raises the bound's $\sigr$ by a tenth as much (the ``Destroys'' column). To understand this behaviour, let us focus on stripe filters first. Along the scan ladder, the rank of the filter and its anisotropy grow together, so that their effects cannot be told apart. On the other hand, the ground template separates them, since it is anisotropic but removes only tens of modes (Sect.~\ref{sec:methods:filters}). At orders 5 and 10, most of the power in the removed subspace sits below the reported band. Both estimators then pay essentially nothing, and the ratio stays at its unfiltered value. Thus, as one may expect, an anisotropic and non-invertible filter costs \pcl nothing as long as it does not act inside the band being reported. As the order grows, the template reaches into the band, and the ratio starts to grow at order 15. At order 25, where almost half of the removed power sits inside the band, it exceeds the ratio of the high-pass filter, which suppresses the same low multipoles isotropically without removing any direction. However, we emphasise that this is geometry-dependent: as soon as we shift to the constant-elevation filters, even the order 10 ground template has a sizeable effect on pure \pcl, as more modes are attenuated along the full inference band. 

\subsubsection{Origin of the gap}

To understand where this gap comes from, recall that the two \pcl pipelines differ in how they separate $E$ from $B$, and therefore in the mask each of them can use (Sect.~\ref{sec:methods:pcl}). Harmonic purification needs an apodised mask, and indeed the pure \pcl optima are always wide tapers. The matrix purification instead removes the $E$ power before \texttt{NaMaster} computes any spectrum, so it needs no taper, and its optima are the binary mask or a taper of at most $5^\circ$. The pure \pcl gap therefore mixes two costs, that of the apodised mask and that of the harmonic $E/B$ separation. To tell them apart, we also evaluate the matrix-purified maps at the mask that the pure \pcl pipeline selects in each state. What the matrix-purified pipeline still loses there is the cost of the mask alone, and it is measured to be approximately half of the gap.

The remaining harmonic part of the gap is the $E \to B$ mixing that the filter has already put in the map, which $B$-purification cannot remove at any resolution, since it corrects only the mixing induced by the known mask (Sect.~\ref{sec:methods:pcl}). The transfer function then corrects the mean of that leak (Sect.~\ref{sec:results:leak}), but not the variance it adds to the $B$ bandpowers, so applying it changes the pure \pcl $\sigr$ by at most $1.4\%$, and by $5.6\%$ under the projector (Appendix~\ref{app:tf:sigr}). The other two methods deal with the leaked $E$ power directly. The matrix purification removes it from the map, while QML weights it through the $E$ and $B$ covariance of the filtered data in $\Cmat^{-1}$, and the two then reach nearly the same error bar.

\subsubsection{Dependence on the noise level}\label{sec:dep_on_noise}

So far we have discussed the reference depth of $2\ukam$. The ordering of the three estimators does not depend on the noise level, but the size of the \pcl gap does. To see this, we first compare the two depths of Table~\ref{tab:sigr_filtered}. Going from $2\ukam$ to $3.5\ukam$ shrinks the pure \pcl gap in every state, and more under the scan filter. This is because part of the pure \pcl excess is the sample variance of the $E$ power that leaks into $B$, which does not grow with the noise, and therefore weighs less as the noise grows. The trend continues at higher noise. On the first realisation set alone, we repeat the unfiltered and the order-3 scan cases at $5$ and $10\ukam$. At $10\ukam$, where the noise dominates the band, the pure \pcl ratio falls to $1.07$ unfiltered and to $1.25$ under the scan filter. Correlated noise acts in the same direction, since the optimistic $1/f$ increment at $2\ukam$ lowers the pure \pcl ratio by a few per cent under the scan filter. The matrix-purified ratio stays within a few per cent of unity in all these cases. Thus, the advantage of QML on $\sigr$ is largest for deep maps, and shrinks significantly as the noise takes over the band.

\subsection{Distance from the pixel-likelihood bound}
\label{sec:results:bound}

\begin{figure*}[t]
  \centering
  \includegraphics[width=\textwidth]{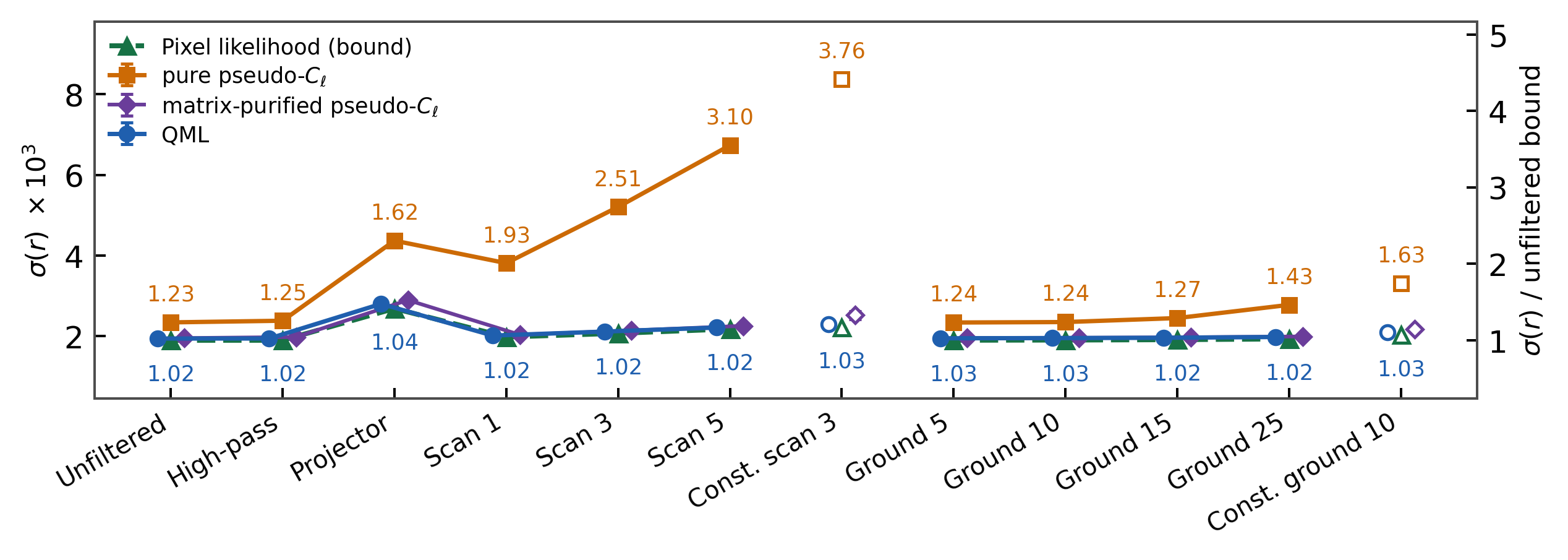}
  \caption{The pixel likelihood as an optimality anchor: $\sigr$ for QML, the two \pcl pipelines and the pixel-likelihood bound of Sect.~\ref{sec:methods:pics} along the filter ladder, all with the $30 \le \ell \le 64$ tensor template. The left axis reports the absolute value of $\sigr$, while the right one is the same quantity in units of the unfiltered bound. Each state is labelled with the pure \pcl $\sigr$ relative to the bound in that state (above), the product of the ``Pure/QML'' and ``QML/bound'' columns of Table~\ref{tab:sigr_filtered}, and with QML's (below). The hollow points are the two constant-elevation states of Table~\ref{tab:sigr_filtered}, distinguished to highlight the underlying different geometry of those cases.}
  \label{fig:picslike_bound}
\end{figure*}

Figure~\ref{fig:picslike_bound} places the three estimators against the pixel-likelihood bound of Sect.~\ref{sec:methods:pics}, the smallest error bar any estimator can reach on these maps, and the ``QML/bound'' column of Table~\ref{tab:sigr_filtered} gives the distance of QML from it.

The bound itself rises with the filter, since a filter that removes modes also removes information. This rise is the ``Destroys'' column of Table~\ref{tab:sigr_filtered}, the irreducible price of each operator, which no estimator can recover. As expected, the high-pass filter and the ground template destroy almost nothing, and the projector destroys the most, since it removes the low multipoles that weigh most on $r$. The scan filter lies in between, and its price grows with the polynomial order. 

Thus, we restate the results of Sect.~\ref{sec:results:ratio}: QML itself sits $2$ to $3\%$ above the bound in every state, and $4\%$ under the projector, while the pure \pcl pipeline pays several times more (Fig.~\ref{fig:picslike_bound}), and most under the scan filter at orders 3 and 5. The matrix-purified pipeline sits $1.5$ to $4.6\%$ above the full bound, most under the projector. The matrix-purified pipeline then sits about as far above its own bound as QML does above the full one. Thus, it reaches QML's $\sigr$ to within the information the purification discards. Under the constant-elevation filters, the results become even more favourable to QML, and the scan case represents the only deviation we could find between matrix-purified and QML (about 10\%).

\subsection{Bandpower errors}
\label{sec:results:bandpowers}

Before the compression onto a template, the estimators can also be compared on the bandpowers themselves, which shows where in the band each filter acts. For this, we use the marginal and conditional errors of Sect.~\ref{sec:methods:protocol}. This is especially important for \pcl bandpowers, which are typically more correlated than QML's (Paper~I).

\begin{figure*}[t]
  \centering
  \includegraphics[width=\textwidth]{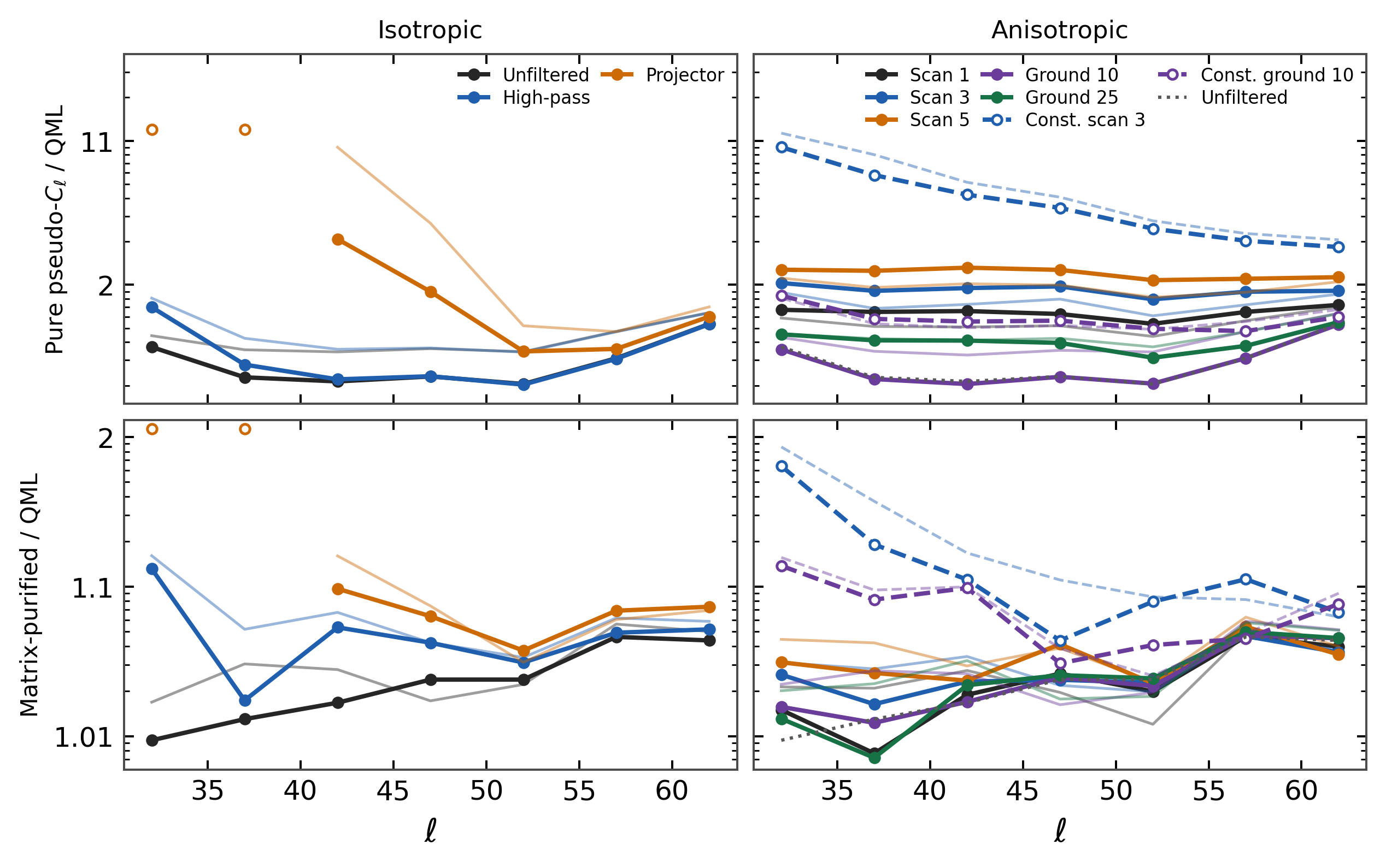}
  \caption{Ratio \pcl/QML of the error on each reported $BB$ bandpower, response-normalised, at $2\ukam$ over the 1000 realisations of the first set. Both \pcl pipelines are first corrected with the $BB$ block of their own transfer matrix (Sect.~\ref{sec:methods:protocol}), which, under the projector, keeps its $\ell \ge 30$ columns, as in Appendix~\ref{app:tf:mean}. Top: pure \pcl. Bottom: matrix-purified \pcl. Dark curves with markers are the conditional error from the full bandpower covariance; light curves are the marginal error from its diagonal. Left: the isotropic states. Right: the anisotropic ones, with the constant-elevation (``Const.'') states of Table~\ref{tab:sigr_filtered} dashed and hollow in the colour of their stripe counterparts, and the unfiltered conditional ratio dotted for reference. The vertical axes are logarithmic in the excess over parity, the ratio minus one, and labelled by the ratio. Under the projector, the two bins adjacent to the nulled subspace, $\ell < 40$, are drawn hollow at the top of the panel, since the filter suppresses both estimators' responses, making them ill-defined.}
  \label{fig:bandpower_ratio}
\end{figure*}

The top row of Fig.~\ref{fig:bandpower_ratio} shows both ratios in every reported bin for the pure \pcl pipeline. Without filtering, the conditional ratio is lowest in the interior of the band and rises at its two edges. The marginal ratio sits above it in every bin, as anticipated. As expected, the high-pass filter raises mainly the lowest bin. The projector acts in the same place with much greater force, and its effect fades above $\ell \simeq 50$. Below $\ell = 40$, the comparison ceases to be meaningful, since the filter has removed most of both estimators' response.

The scan filter behaves differently. Being anisotropic, it acts at every multipole of the band, and indeed it raises the ratio in every bin at once, and further at each order, with no return to the unfiltered value anywhere in the band. At orders 3 and 5, the largest excess sits at the lowest multipoles. The ground template at order 10 lies on top of the unfiltered curve in every bin, while at order 25 it raises the lower half of the band. Finally, under the scan filter, the marginal ratio falls below the conditional one, the reverse of the other states. There, the conjugated QML bandpowers are more correlated than the pure \pcl ones. The constant-elevation filters raise the ratio well above the stripe filters of the same order (Fig.~\ref{fig:bandpower_ratio}, dashed). Under the scan filter it falls from $10.0$ in the lowest bin to $2.8$ in the highest, and under the ground filter it sits at $1.5$ to $1.8$, above the unfiltered curve in every bin.

The bottom row of Fig.~\ref{fig:bandpower_ratio} shows the matrix-purified pipeline, corrected like the pure \pcl one with its own transfer matrix. There, the conditional ratios sit at $1.01$ to $1.05$ in every bin of every state outside the constant-elevation filters, except the lowest bin under the high-pass filter, at $1.13$, and the bins above $\ell = 40$ under the projector, at $1.04$ to $1.10$. Under the constant-elevation filters, they reach $1.64$ in the lowest bin under the scan filter and $1.14$ under the ground filter. This is in line with its $\sigr$ ratios in Table~\ref{tab:sigr_filtered}.

\subsection{$E \to B$ leakage and the matrix transfer function}
\label{sec:results:leak}

We finally turn from the error bars to the bias, looking at the $E \to B$ leakage, that is, the $BB$ power that a sky with only $E$ modes produces in each estimator. It matters because the lensing $BB$ signal is small, so that even a small fraction of the $E$ power leaking into $B$ can bias the $BB$ bandpowers. Table~\ref{tab:leak} gives the measured leakage of QML and of the two \pcl pipelines, averaged over the band as a fraction of the lensing $BB$ signal (Sect.~\ref{sec:methods:protocol}). For the pure \pcl pipeline, it reports the leak before and after the matrix transfer function of Eq.~\eqref{eq:tf}, measured at the same configuration. Without the correction, the pure \pcl leak is a sizeable fraction of the lensing $BB$ signal in every state, and exceeds it under the projector. Outside the projector, this is two to three orders of magnitude above QML's leak.

\begin{table}[!htb]
  \centering \footnotesize \setlength{\tabcolsep}{3pt}
  \caption{$E\to B$ leakage relative to the $BB$ lensing signal.}
  \label{tab:leak}
  \begin{tabular}{lcccccc}
    \toprule
    & \multicolumn{2}{c}{QML} & \multicolumn{3}{c}{Pure \pcl} & Matrix \\
    \cmidrule(lr){2-3}\cmidrule(lr){4-6}
    State & Leak & Floor & Leak & $+\,\Tmat$ & Lowest & $\times 10^{4}$ \\
    \midrule
    Unfiltered    & 0.0010 & 0.0001 & 0.18 & 0.015 & 0.06 & 2.5 \\
    High-pass     & 0.0010 & 0.0001 & 0.22 & 0.033 & 0.07 & 2.5 \\
    Projector     & 0.12   & 0.04   & 2.93 & 0.6   & 2.06 & 10.0 \\
    Scan 1        & 0.0011 & 0.0001 & 0.50 & 0.018 & 0.37 & 2.6 \\
    Scan 3        & 0.0012 & 0.0001 & 0.61 & 0.006 & 0.52 & 3.0 \\
    Scan 5        & 0.0012 & 0.0002 & 0.83 & 0.007 & 0.76 & 3.3 \\
    Const. scan 3 & 0.0027 & 0.0002 & 2.80 & 0.160 & 2.36 & 3.1 \\
    \bottomrule
  \end{tabular}
  \tablefoot{Values are at $2\ukam$ over the reported band, on the first realisation set, with the leakage defined as in Sect.~\ref{sec:methods:protocol}. For QML, the first column shows the amount we measure and the second its Monte Carlo noise. The pure \pcl columns give the leak before and after the matrix transfer function of Eq.~\eqref{eq:tf}, at the configuration with the lowest $\sigr$ averaged over the three realisation sets, and ``Lowest'' is the smallest uncorrected leak any of the 20 configurations reaches. The Monte Carlo errors of the corrected leak are in Appendix~\ref{app:tf:mean}. Under the projector, the transfer matrix is built on its $\ell \ge 30$ columns only, and its corrected leak is dominated by Monte Carlo noise. QML's entry there is set by the ill-conditioned buffer bins (see text). The last column is the matrix-purified pipeline, selected in the same way, in units of $10^{-4}$. Const. scan 3 is as in Table~\ref{tab:sigr_filtered}.}
\end{table}

Two properties of this leak matter for an analysis. First, it is an additive bias, which the decoupling of Eq.~\eqref{eq:master} does not remove. Indeed, it lives off the diagonal of the transfer matrix, in the terms that map $EE$ into $BB$, so it must be measured from simulations and then subtracted or marginalised. This is what the matrix transfer function does. It removes most of the leak, and at scan orders 3 and 5 what remains is consistent with zero (Table~\ref{tab:leak}). Under the constant-elevation scan filter, instead, the uncorrected leak is almost three times the lensing $BB$ signal, and after the correction it is still about a sixth of it, three times its Monte Carlo error. Under the projector, the transfer function must leave out its $20 \le \ell \le 29$ columns, which probe the nulled subspace (Appendix~\ref{app:tf:mean}), and the corrected leak is then dominated by Monte Carlo noise. A clearer measure is the bias that the mean leak puts on $r$ (Sect.~\ref{sec:methods:protocol}), since this is what an analysis of $r$ would actually suffer. On the first realisation set, before the correction it is $1.8$ to $2.7\,\sigr$ unfiltered and under the isotropic and scan filters, $5.9\,\sigr$ under the constant-elevation scan filter, and after it at most $0.13\,\sigr$ (Appendix~\ref{app:tf:mean}). Under the constant-elevation ground filter, not in Table~\ref{tab:leak}, it goes from $3.8$ to $0.30\,\sigr$.

Furthermore, the leak keeps falling with apodisation long after the error bar has started to worsen (Appendix~\ref{app:unfiltered:mask}). Thus, the pure \pcl configuration with the lowest leak is not the one with the best $\sigr$, and choosing it costs up to $23\%$ on $\sigr$.

QML's own leak is $10^{-3}$ of the lensing signal, with no knobs at all. This is well above its Monte Carlo floor, so it is small but measured. It grows without the low-side buffer, as expected if it comes from $E$ power outside the estimated range, which enters the reported bins through the out-of-band term of Eq.~\eqref{eq:qml_y}. Under the projector, QML's leak is much larger and varies between realisation sets from $0.015$ to $0.12$. This is a numerical effect, since the $20 \le \ell \le 29$ buffer bins sit in the nulled subspace (Appendix~\ref{app:checks:buffer}). Indeed, without them, QML's leak returns to the level of the other states. QML's leak moves $r$ by at most $0.01\,\sigr$ unfiltered and under the isotropic and scan filters, and by at most $0.015\,\sigr$ under the constant-elevation filters. The matrix-purified pipeline has the lowest leak of the three, below QML's in every state, but again it is measured to be above its Monte Carlo noise.

Beyond the size of the leak, the pipelines differ in what they need to control it. Our delta-bin transfer function uses $30\,000$ signal-only simulations for each pure \pcl configuration and each filter. QML needs no transfer function, and the matrix-purified pipeline needs one only to calibrate its $B$ response (Sect.~\ref{sec:methods:pcl}), but both need the filter as a matrix.

\section{Conclusions}
\label{sec:conclusions}

In this work, we have compared the QML estimator implemented in \cosmoforge (\qube) with two \pcl power spectrum pipelines. We considered data filtered by a ladder of isotropic and anisotropic filters, the latter mimicking TOD filtering built as a constant-elevation scan at the South Pole or at Atacama. As regards QML, the filters enter the estimator as exact pixel-space operators. Instead, the \pcl side can be divided into two approaches: the first is the pure \pcl estimator, with harmonic $B$-purification; the second separates $E$ from $B$ in pixel space before the spectra are computed, with a matrix purification built from the same filter and fiducial spectra as QML. Both \pcl pipelines need to be corrected afterwards by a transfer function that allows us to restore the power lost to the filters. In addition, the results of these pipelines are compared with a pixel-based likelihood (\pics, Paper~I), which represents our ground truth. Our conclusions are the following.

\begin{enumerate}
\item With no knobs to tune, QML sits $2$ to $3\%$ above the pixel-likelihood bound on $\sigr$ in every filter state, and $4\%$ under the projector, whose low-side buffer bins sit in the nulled subspace. Relative to its own unfiltered value, its $\sigr$ grows following the amount of information removed by the filter, from under $0.1\%$ under a high-pass filter to $44\%$ under an $\ell < 30$ projector.
\item Unfiltered and at $2\ukam$, the pure \pcl pipeline trails QML by $21\%$ on $\sigr$ at its tuned optimum, and an untuned configuration could be several times worse. At $N_{\rm side} = 64$, this gap falls to $10\%$, mostly from a better-resolved harmonic purification at a fixed taper and partly from the narrower taper that the finer pixels make affordable (Appendix~\ref{app:checks:binning}). Filtering opens the gap, moderately under the isotropic filters and most under the scan filters. Along stripes of constant declination, the $\sigr$ ratio reaches $3.0$ at polynomial order 5. Under the order-3 scan filter rebuilt from a constant-elevation scan, it reaches $3.7$. A low-rank ground template leaves the ratio unchanged as long as it acts below the reported band, whereas the order-10 ground filter rebuilt from the same scan, which attenuates modes across the band, raises it to $1.6$.
\item The matrix-purified $\sigr$ is $1.00$ to $1.03$ times QML's in almost all states. In these states, the gap of item 2 is the cost of separating $E$ from $B$ harmonically after the filter, together with the apodisation this requires. A transfer function does not remove the residual, since applying one changes the pure \pcl $\sigr$ by at most $1.4\%$, and by $5.6\%$ under the projector. The matrix purification needs the filter operator, similarly to the QML, and the $E$ and $B$ signal covariances of the filtered data that \qube provides, but its bandpowers are biased low and need a transfer function of their own.
\item QML's measured $E \to B$ leakage is $10^{-3}$ of the lensing $BB$ signal outside the projector, $3\times10^{-3}$ under the constant-elevation scan filter, and moves $r$ by at most $0.01\,\sigr$ and by at most $0.015\,\sigr$ under the constant-elevation filters. The matrix-purified pipeline's leak is a few times lower. Before its transfer function, the pure \pcl leak is enough to bias $r$ by $1.8$ to $2.7\,\sigr$ unfiltered and under the isotropic and scan filters, and by $5.9\,\sigr$ under the constant-elevation scan filter. After correcting for the transfer function, the bias is reduced to at most $0.13\,\sigr$ in every state of Table~\ref{tab:leak}, but leaves $0.30\,\sigr$ under the constant-elevation ground filter. When dealing with the projector, the nulled scales need to be left out of the correction as they are ill-defined.
\item The size of these gaps depends on the noise level. The pure \pcl gap narrows as the noise grows, so comparing the results for $2\ukam$ to the ones where noise dominates at $10\ukam$, the degradation goes from 21\% to $7\%$ without filtering, and from a factor of 2.46 to $25\%$ under the order-3 scan filter. At $2\ukam$, the optimistic $1/f$ increment lowers the ratio under the scan filter by a few per cent, but that is expected to have a stronger effect for higher noise levels, or less optimistic assumptions on the $1/f$ knee. Thus, for noisier maps we expect the gap to keep narrowing, and the advantage of QML to rest increasingly on the absence of tuning knobs and of transfer-function simulations.
\end{enumerate}

The next step is to apply \cosmoforge to a real SAT observation matrix, in which the TOD filter and the map-maker are composed into one linear operator on the map vector, and to compare the results with both \pcl pipelines on the same data. Such an application would also lift the idealisations of the present setup. For example, the scan-filter rows describe a single-direction stripe filter acting on $Q$ and $U$ separately in the equatorial frame, without the rotation of the instrument frame that mixes them for data demodulated by a rotating half-wave plate \citep{SOforecast, HWPmodulation}, and without the cross-linking of a real scan strategy. Each stripe is also a whole ring segment of the footprint, longer than a typical subscan, one constant-elevation sweep in a single azimuth direction \citep{SOtransfer}, so our polynomial orders do not map onto those of a TOD pipeline. A real pipeline fits and subtracts a polynomial per scan and then averages residuals from several parallactic angles, that is, orientations of the instrument on the sky, into the same pixel, so the composite operator is a weighted average of the per-angle filters rather than their union. Likewise, a real ground template is fitted per observation, so the map-space operator is an average of many templates rather than the single global template we remove, and our order is not comparable to the per-observation azimuthal order of a real pipeline. The constant-elevation filters of Sect.~\ref{sec:methods:filters} are such operators, for a toy scan strategy at an Atacama-like site. They attenuate most modes rather than removing them and, with the noise filtered as the time streams are, they widen the gap between pure \pcl and QML, while the matrix-purified pipeline sits a few to ten per cent above QML. Still, the filter we built is not representative of a fully realistic analysis pipeline.

Also, QML and the matrix purification are both handed the filter exactly, as a matrix, together with the true fiducial spectra, and QML also the exact noise covariance the noise was drawn from, including the $1/f$ term, which here is isotropic rather than striped along the scan. A wrong fiducial would cost QML optimality but would not bias it \citep{Tegmark1997}. For the purification, it would change which directions are kept, and with them the small leak of Sect.~\ref{sec:results:leak}, although the construction works even with the generic $C_\ell \propto \ell^{-2}$ spectra of \citet{ObsMatrix}. A mismatch between the operator used to build $\Rmat$ and the one applied to the data degrades both, and by how much is a question for a real pipeline application. Also, here we consider only noise without any foreground, whose mitigation is very important in a real-life scenario. Finally, the pure \pcl grid uses standard apodisation kernels only, and optimised pure-$B$ windows \citep{SmithZaldarriaga2007, Grain2009, Ferte2013} could lower its error bars, in particular at the lowest multipoles.

Despite this, \cosmoforge provides, from a single model of the filtered data, both the QML bandpowers and the pixel-likelihood bound that measures how close they come to optimal. We showed that this replaces the configuration search and the transfer-function simulations of a \pcl analysis, at the price of handing the estimator the filter as a matrix. In addition, \cosmoforge provides the main ingredients to perform the matrix-purified \pcl. 

Finally, we emphasise that the SO SAT footprint and noise levels serve here only as a proxy for a generic ground-based experiment. Thus, the results of this work are not a statement on present or future SO forecasts.

\section*{Data availability}
The new filter interface is part of the public \cosmoforge release 1.3.0, so the comparison in Sect.~\ref{sec:results} can be repeated with any linear operator on the map vector that a pipeline provides.

\begin{acknowledgements}
The author thanks Magdy Morshed and Luca Pagano for valuable comments and discussions. Some of the results in this paper have been derived using the following packages: {\tt CAMB} \citep{Lewis2000, Howlett2012}, {\tt healpy} \citep{Gorski2005, Zonca2019}, {\tt NaMaster} \citep{Alonso2019}, {\tt Matplotlib} \citep{Hunter2007}, {\tt SciPy} \citep{Virtanen2020} and {\tt NumPy} \citep{Harris2020}. We acknowledge the financial support from the INFN InDark initiative and from the COSMOS network through the ASI (Italian Space Agency) Grants 2016-24-H.0 and 2016-24-H.1-2018. This work has also received funding by the European Union's Horizon 2020 research and innovation programme under grant agreement no.\ 101007633 CMB-Inflate. We acknowledge support from the MUR PRIN2022 Project ``BROWSEPOL: Beyond standaRd mOdel With coSmic microwavE background POLarization''-2022EJNZ53 financed by the European Union -- Next Generation EU. We acknowledge CINECA for the availability of high-performance computing resources and support through the CINECA-INFN agreement. During the preparation of this work, the author used Anthropic's Claude (via the Claude Code agentic coding interface) to assist with code implementation and manuscript drafting. The author reviewed, edited, and validated all AI-generated content and code and takes full responsibility for the contents of this publication.
\end{acknowledgements}

\bibliographystyle{bibtex/aa}
\bibliography{references}

@article{SOforecast,
  author  = {{Ade}, P. and others},
  collaboration = {Simons Observatory},
  title   = {The Simons Observatory: science goals and forecasts},
  journal = {JCAP},
  volume  = {02},
  pages   = {056},
  year    = {2019},
  doi     = {10.1088/1475-7516/2019/02/056},
  eprint  = {1808.07445}
}

@article{Tegmark1997,
  author  = {{Tegmark}, M.},
  title   = {How to measure CMB power spectra without losing information},
  journal = {Phys. Rev. D},
  volume  = {55},
  pages   = {5895},
  year    = {1997},
  doi     = {10.1103/PhysRevD.55.5895},
  eprint  = {astro-ph/9611174}
}

@article{TegmarkOliveira2001,
  author  = {{Tegmark}, M. and {de Oliveira-Costa}, A.},
  title   = {How to measure CMB polarization power spectra without losing information},
  journal = {Phys. Rev. D},
  volume  = {64},
  pages   = {063001},
  year    = {2001},
  doi     = {10.1103/PhysRevD.64.063001},
  eprint  = {astro-ph/0012120}
}

@article{TTH1997,
  author  = {{Tegmark}, M. and {Taylor}, A.~N. and {Heavens}, A.~F.},
  title   = {Karhunen-Loeve Eigenvalue Problems in Cosmology: How Should We Tackle Large Data Sets?},
  journal = {ApJ},
  volume  = {480},
  pages   = {22},
  year    = {1997},
  doi     = {10.1086/303939},
  eprint  = {astro-ph/9603021}
}

@article{MASTER,
  author  = {{Hivon}, E. and {G{\'o}rski}, K.~M. and {Netterfield}, C.~B. and {Crill}, B.~P. and {Prunet}, S. and {Hansen}, F.},
  title   = {MASTER of the Cosmic Microwave Background Anisotropy Power Spectrum: A Fast Method for Statistical Analysis of Large and Complex Cosmic Microwave Background Data Sets},
  journal = {ApJ},
  volume  = {567},
  pages   = {2},
  year    = {2002},
  doi     = {10.1086/338126},
  eprint  = {astro-ph/0105302}
}

@article{BK18,
  author  = {{Ade}, P.~A.~R. and others},
  collaboration = {BICEP/Keck},
  title   = {Improved Constraints on Primordial Gravitational Waves using Planck, WMAP, and BICEP/Keck Observations through the 2018 Observing Season},
  journal = {Phys. Rev. Lett.},
  volume  = {127},
  pages   = {151301},
  year    = {2021},
  doi     = {10.1103/PhysRevLett.127.151301},
  eprint  = {2110.00483}
}

@article{Smith2006,
  author  = {{Smith}, K.~M.},
  title   = {Pseudo-$C_\ell$ estimators which do not mix E and B modes},
  journal = {Phys. Rev. D},
  volume  = {74},
  pages   = {083002},
  year    = {2006},
  doi     = {10.1103/PhysRevD.74.083002},
  eprint  = {astro-ph/0511629}
}

@article{Grain2009,
  author  = {{Grain}, J. and {Tristram}, M. and {Stompor}, R.},
  title   = {Polarized CMB power spectrum estimation using the pure pseudo-cross-spectrum approach},
  journal = {Phys. Rev. D},
  volume  = {79},
  pages   = {123515},
  year    = {2009},
  doi     = {10.1103/PhysRevD.79.123515},
  eprint  = {0903.2350}
}

@article{Wolz2024,
  author  = {{Wolz}, K. and others},
  collaboration = {Simons Observatory},
  title   = {The Simons Observatory: Pipeline comparison and validation for large-scale B-modes},
  journal = {A\&A},
  volume  = {686},
  pages   = {A16},
  year    = {2024},
  doi     = {10.1051/0004-6361/202346105},
  eprint  = {2302.04276}
}

@article{SOtransfer,
  author  = {{Herv{\'i}as-Caimapo}, C. and {Wolz}, K. and {La Posta}, A. and others},
  collaboration = {Simons Observatory},
  title   = {The Simons Observatory: Validation of reconstructed power spectra from simulated filtered maps for the Small Aperture Telescope survey},
  journal = {JCAP},
  volume  = {06},
  pages   = {055},
  year    = {2025},
  doi     = {10.1088/1475-7516/2025/06/055},
  eprint  = {2502.00946}
}

@article{SOsat2025,
  author  = {{Abril-Cabezas}, I. and others},
  collaboration = {Simons Observatory},
  title   = {The Simons Observatory: forecasted constraints on primordial gravitational waves with the expanded array of Small Aperture Telescopes},
  journal = {JCAP},
  volume  = {04},
  pages   = {051},
  year    = {2026},
  doi     = {10.1088/1475-7516/2026/04/051},
  eprint  = {2512.15833}
}

@article{Hartlap2007,
  author  = {{Hartlap}, J. and {Simon}, P. and {Schneider}, P.},
  title   = {Why your model parameter confidences might be too optimistic. Unbiased estimation of the inverse covariance matrix},
  journal = {A\&A},
  volume  = {464},
  pages   = {399},
  year    = {2007},
  doi     = {10.1051/0004-6361:20066170},
  eprint  = {astro-ph/0608064}
}

@article{CosmoForgeI,
  author  = {{Galloni}, G. and {Pagano}, L.},
  title   = {CosmoForge I: A unified framework for QML power spectrum estimation and pixel-based likelihood analysis},
  journal = {A\&A},
  year    = {2026},
  note    = {in press},
  doi     = {10.48550/arXiv.2605.21149},
  eprint  = {2605.21149}
}

@article{BondJaffeKnox1998,
  author  = {{Bond}, J.~R. and {Jaffe}, A.~H. and {Knox}, L.},
  title   = {Estimating the power spectrum of the cosmic microwave background},
  journal = {Phys. Rev. D},
  volume  = {57},
  pages   = {2117},
  year    = {1998},
  doi     = {10.1103/PhysRevD.57.2117},
  eprint  = {astro-ph/9708203}
}

@article{TODfiltering,
  author  = {{Ade}, P.~A.~R. and others},
  collaboration = {BICEP2},
  title   = {Detection of B-Mode Polarization at Degree Angular Scales by BICEP2},
  journal = {Phys. Rev. Lett.},
  volume  = {112},
  pages   = {241101},
  year    = {2014},
  doi     = {10.1103/PhysRevLett.112.241101},
  eprint  = {1403.3985}
}

@article{BICEP1filtering,
  author  = {{Chiang}, H.~C. and others},
  collaboration = {BICEP},
  title   = {Measurement of Cosmic Microwave Background Polarization Power Spectra from Two Years of BICEP Data},
  journal = {ApJ},
  volume  = {711},
  pages   = {1123},
  year    = {2010},
  doi     = {10.1088/0004-637X/711/2/1123},
  eprint  = {0906.1181}
}

@article{BICEP1threeyear,
  author  = {{Barkats}, D. and others},
  collaboration = {BICEP1},
  title   = {Degree-scale Cosmic Microwave Background Polarization Measurements from Three Years of BICEP1 Data},
  journal = {ApJ},
  volume  = {783},
  pages   = {67},
  year    = {2014},
  doi     = {10.1088/0004-637X/783/2/67},
  eprint  = {1310.1422}
}

@article{SPTpolBB,
  author  = {{Sayre}, J.~T. and others},
  collaboration = {SPTpol},
  title   = {Measurements of B-mode polarization of the cosmic microwave background from 500 square degrees of SPTpol data},
  journal = {Phys. Rev. D},
  volume  = {101},
  pages   = {122003},
  year    = {2020},
  doi     = {10.1103/PhysRevD.101.122003},
  eprint  = {1910.05748}
}

@article{HWPmodulation,
  author  = {{Kusaka}, A. and others},
  title   = {Results from the Atacama B-mode Search (ABS) experiment},
  journal = {JCAP},
  volume  = {09},
  pages   = {005},
  year    = {2018},
  doi     = {10.1088/1475-7516/2018/09/005},
  eprint  = {1801.01218}
}

@article{ObsMatrix,
  author  = {{Ade}, P.~A.~R. and others},
  collaboration = {Keck Array and BICEP2},
  title   = {BICEP2/Keck Array. VII. Matrix Based E/B Separation Applied to BICEP2 and the Keck Array},
  journal = {ApJ},
  volume  = {825},
  pages   = {66},
  year    = {2016},
  doi     = {10.3847/0004-637X/825/1/66},
  eprint  = {1603.05976}
}

@article{BKXX,
  author  = {{Ade}, P.~A.~R. and others},
  collaboration = {BICEP/Keck},
  title   = {BICEP/Keck. XX. Component-separated Maps of the Polarized Cosmic Microwave Background and Thermal Dust Emission Using Planck and BICEP/Keck Observations through the 2018 Observing Season},
  journal = {ApJ},
  volume  = {999},
  pages   = {89},
  year    = {2026},
  doi     = {10.3847/1538-4357/ae3e53},
  eprint  = {2509.21648}
}

@article{BICEP2instrument,
  author  = {{Ade}, P.~A.~R. and others},
  collaboration = {BICEP2},
  title   = {BICEP2. II. Experiment and three-year Data Set},
  journal = {Astrophys. J.},
  volume  = {792},
  pages   = {62},
  year    = {2014},
  doi     = {10.1088/0004-637X/792/1/62},
  eprint  = {1403.4302}
}

@article{Delensing2021,
  author  = {{Ade}, P.~A.~R. and others},
  collaboration = {BICEP/Keck and SPTpol},
  title   = {A demonstration of improved constraints on primordial gravitational waves with delensing},
  journal = {Phys. Rev. D},
  volume  = {103},
  pages   = {022004},
  year    = {2021},
  doi     = {10.1103/PhysRevD.103.022004},
  eprint  = {2011.08163}
}

@ARTICLE{Lewis2000,
    author  = {Lewis, A. and Challinor, A. and Lasenby, A.},
    title   = {Efficient Computation of Cosmic Microwave Background Anisotropies in Closed {Friedmann-Robertson-Walker} Models},
    journal = {ApJ},
    volume  = {538},
    pages   = {473},
    year    = {2000},
    doi     = {10.1086/309179},
    eprint  = {astro-ph/9911177},
    archivePrefix = {arXiv}
}

@ARTICLE{Howlett2012,
    author  = {Howlett, C. and Lewis, A. and Hall, A. and Challinor, A.},
    title   = {{CMB} power spectrum parameter degeneracies in the era of precision cosmology},
    journal = {JCAP},
    volume  = {2012},
    number  = {04},
    pages   = {027},
    year    = {2012},
    doi     = {10.1088/1475-7516/2012/04/027},
    eprint  = {1201.3654},
    archivePrefix = {arXiv}
}

@ARTICLE{Gorski2005,
    author  = {G{\'o}rski, K. M. and Hivon, E. and Banday, A. J. and Wandelt, B. D. and Hansen, F. K. and Reinecke, M. and Bartelmann, M.},
    title   = {{HEALPix}: a framework for high-resolution discretization and fast analysis of data distributed on the sphere},
    journal = {ApJ},
    volume  = {622},
    pages   = {759},
    year    = {2005}
}

@ARTICLE{Zonca2019,
    author  = {Zonca, A. and Singer, L. and Lenz, D. and Reinecke, M. and Rosset, C. and Hivon, E. and Gorski, K.},
    title   = {{healpy}: equal area pixelization and spherical harmonics transforms for data on the sphere in {Python}},
    journal = {J.\ Open Source Softw.},
    volume  = {4},
    pages   = {1298},
    year    = {2019}
}

@ARTICLE{Alonso2019,
    author  = {Alonso, D. and Sanchez, J. and Slosar, A. and {LSST Dark Energy Science Collaboration}},
    title   = {A unified pseudo-{$C_\ell$} framework},
    journal = {MNRAS},
    volume  = {484},
    pages   = {4127},
    year    = {2019},
    eprint  = {1809.09603},
    archivePrefix = {arXiv},
    doi     = {10.1093/mnras/stz093}
}

@ARTICLE{Hunter2007,
    author  = {Hunter, J. D.},
    title   = {{Matplotlib}: a {2D} graphics environment},
    journal = {Comput.\ Sci.\ Eng.},
    volume  = {9},
    pages   = {90},
    year    = {2007}
}

@ARTICLE{Virtanen2020,
    author  = {Virtanen, P. and Gommers, R. and Oliphant, T. E. and others},
    title   = {{SciPy} 1.0: fundamental algorithms for scientific computing in {Python}},
    journal = {Nat.\ Methods},
    volume  = {17},
    pages   = {261},
    year    = {2020}
}

@ARTICLE{Harris2020,
    author  = {Harris, C. R. and Millman, K. J. and van der Walt, S. J. and others},
    title   = {Array programming with {NumPy}},
    journal = {Nature},
    volume  = {585},
    pages   = {357},
    year    = {2020}
}

@inproceedings{LiteBIRDoverview,
    author = "Aizawa, K. and others",
    collaboration = "LiteBIRD",
    title = "{LiteBIRD Mission Overview after Mission Reformation}",
    booktitle = "{SPIE Astronomical Telescopes + Instrumentation 2026}",
    eprint = "2607.25285",
    archivePrefix = "arXiv",
    primaryClass = "astro-ph.IM",
    month = "7",
    year = "2026",
  note = {[arXiv:2607.25285]}
}

@article{Planck2018VI,
    author = "Aghanim, N. and others",
    collaboration = "Planck",
    title = "{Planck 2018 results. VI. Cosmological parameters}",
    eprint = "1807.06209",
    archivePrefix = "arXiv",
    primaryClass = "astro-ph.CO",
    doi = "10.1051/0004-6361/201833910",
    journal = {A\&A},
    volume = "641",
    pages = "A6",
    year = "2020",
}

@article{WMAP9,
    author = "Bennett, C. L. and others",
    collaboration = "WMAP",
    title = "{Nine-Year Wilkinson Microwave Anisotropy Probe (WMAP) Observations: Final Maps and Results}",
    eprint = "1212.5225",
    archivePrefix = "arXiv",
    primaryClass = "astro-ph.CO",
    doi = "10.1088/0067-0049/208/2/20",
    journal = {ApJS},
    volume = "208",
    pages = "20",
    year = "2013"
}

@article{Galloni2024,
    author = "Galloni, Giacomo and Henrot-Versill\'e, Sophie and Tristram, Matthieu",
    title = "{Robust constraints on tensor perturbations from cosmological data: A comparative analysis from Bayesian and frequentist perspectives}",
    eprint = "2405.04455",
    archivePrefix = "arXiv",
    primaryClass = "astro-ph.CO",
    doi = "10.1103/PhysRevD.110.063511",
    journal = "Phys. Rev. D",
    volume = "110",
    number = "6",
    pages = "063511",
    year = "2024",
}

@article{Galloni2023,
    author = "Galloni, Giacomo and Bartolo, Nicola and Matarrese, Sabino and Migliaccio, Marina and Ricciardone, Angelo and Vittorio, Nicola",
    title = "{Updated constraints on amplitude and tilt of the tensor primordial spectrum}",
    eprint = "2208.00188",
    archivePrefix = "arXiv",
    primaryClass = "astro-ph.CO",
    doi = "10.1088/1475-7516/2023/04/062",
    journal = "JCAP",
    volume = "04",
    pages = "062",
    year = "2023",
}

@article{LiteBIRDinflation,
  title = {Probing {{Cosmic Inflation}} with the {{LiteBIRD Cosmic Microwave Background Polarization Survey}}},
  author = {Allys, E. and Arnold, K. and Aumont, J. and Aurlien, R. and Azzoni, S. and Baccigalupi, C. and Banday, A. J. and Banerji, R. and Barreiro, R. B. and Bartolo, N. and Bautista, L. and Beck, D. and Beckman, S. and Bersanelli, M. and Boulanger, F. and Brilenkov, M. and Bucher, M. and Calabrese, E. and Campeti, P. and Carones, A. and Casas, F. J. and Catalano, A. and Chan, V. and Cheung, K. and Chinone, Y. and Clark, S. E. and Columbro, F. and D'Alessandro, G. and {de Bernardis}, P. and {de Haan}, T. and {de la Hoz}, E. and De Petris, M. and Della Torre, S. and {Diego-Palazuelos}, P. and Dobbs, M. and Dotani, T. and Duval, J. M. and Elleflot, T. and Eriksen, H. K. and Errard, J. and {Essinger-Hileman}, T. and Finelli, F. and Flauger, R. and Franceschet, C. and Fuskeland, U. and Galloway, M. and Ganga, K. and Gerbino, M. and Gervasi, M. and {G{\'e}nova-Santos}, R. T. and Ghigna, T. and Giardiello, S. and Gjerl{\o}w, E. and Grain, J. and Grupp, F. and Gruppuso, A. and Gudmundsson, J. E. and Halverson, N. W. and Hargrave, P. and Hasebe, T. and Hasegawa, M. and Hazumi, M. and {Henrot-Versill{\'e}}, S. and Hensley, B. and Hergt, L. T. and Herman, D. and Hivon, E. and Hlozek, R. A. and Hornsby, A. L. and Hoshino, Y. and Hubmayr, J. and Ichiki, K. and Iida, T. and Imada, H. and Ishino, H. and Jaehnig, G. and Katayama, N. and Kato, A. and Keskitalo, R. and Kisner, T. and Kobayashi, Y. and Kogut, A. and Kohri, K. and Komatsu, E. and Komatsu, K. and Konishi, K. and Krachmalnicoff, N. and Kuo, C. L. and Lamagna, L. and Lattanzi, M. and Lee, A. T. and Leloup, C. and Levrier, F. and Linder, E. and Luzzi, G. and {Macias-Perez}, J. and Maciaszek, T. and Maffei, B. and Maino, D. and Mandelli, S. and {Mart{\'i}nez-Gonz{\'a}lez}, E. and Masi, S. and Massa, M. and Matarrese, S. and Matsuda, F. T. and Matsumura, T. and Mele, L. and Migliaccio, M. and Minami, Y. and Moggi, A. and Montgomery, J. and Montier, L. and Morgante, G. and Mot, B. and Nagano, Y. and Nagasaki, T. and Nagata, R. and Nakano, R. and Namikawa, T. and Nati, F. and Natoli, P. and Nerval, S. and Noviello, F. and Odagiri, K. and Oguri, S. and Ohsaki, H. and Pagano, L. and Paiella, A. and Paoletti, D. and Passerini, A. and Patanchon, G. and Piacentini, F. and Piat, M. and Polenta, G. and Poletti, D. and Prouv{\'e}, T. and Puglisi, G. and Rambaud, D. and Raum, C. and Realini, S. and Reinecke, M. and Remazeilles, M. and Ritacco, A. and Roudil, G. and {Rubino-Martin}, J. A. and Russell, M. and Sakurai, H. and Sakurai, Y. and Sasaki, M. and Scott, D. and Sekimoto, Y. and Shinozaki, K. and Shiraishi, M. and Shirron, P. and Signorelli, G. and Spinella, F. and Stever, S. and Stompor, R. and Sugiyama, S. and Sullivan, R. M. and Suzuki, A. and Svalheim, T. L. and Switzer, E. and Takaku, R. and Takakura, H. and Takase, Y. and Tartari, A. and Terao, Y. and Thermeau, J. and Thommesen, H. and Thompson, K. L. and Tomasi, M. and Tominaga, M. and Tristram, M. and Tsuji, M. and Tsujimoto, M. and Vacher, L. and Vielva, P. and Vittorio, N. and Wang, W. and Watanuki, K. and Wehus, I. K. and Weller, J. and Westbrook, B. and Wilms, J. and Wollack, E. J. and Yumoto, J. and Zannoni, M.},
  year = {2023},
  month = nov,
  journal = {PTEP},
  eprint = {2202.02773},
  primaryclass = {astro-ph},
  pages = {042F01},
  issn = {2050-3911},
  doi = {10.1093/ptep/ptac150},
  visitdate = {2022-12-21},
  archiveprefix = {arxiv},
  volume = {2023}
}

@article{Tristram2022,
  ids = {Tristram:2022},
  title = {Improved Limits on the Tensor-to-Scalar Ratio Using {{BICEP}} and {{Planck}} Data},
  author = {Tristram, M. and others},
  year = {2022},
  journal = {Phys. Rev. D},
  volume = {105},
  number = {8},
  eprint = {2112.07961},
  primaryclass = {astro-ph.CO},
  pages = {083524},
  doi = {10.1103/PhysRevD.105.083524},
  archiveprefix = {arxiv}
}

@article{Nicola2021,
  author  = {{Nicola}, A. and {Garc{\'i}a-Garc{\'i}a}, C. and {Alonso}, D. and {Dunkley}, J. and {Ferreira}, P.~G. and {Slosar}, A. and {Spergel}, D.~N.},
  title   = {Cosmic shear power spectra in practice},
  journal = {JCAP},
  volume  = {03},
  pages   = {067},
  year    = {2021},
  doi     = {10.1088/1475-7516/2021/03/067},
  eprint  = {2010.09717}
}

@article{Lewis2002,
  author  = {{Lewis}, A. and {Challinor}, A. and {Turok}, N.},
  title   = {Analysis of CMB polarization on an incomplete sky},
  journal = {Phys. Rev. D},
  volume  = {65},
  pages   = {023505},
  year    = {2002},
  doi     = {10.1103/PhysRevD.65.023505},
  eprint  = {astro-ph/0106536}
}

@article{Bunn2003,
  author  = {{Bunn}, E.~F. and {Zaldarriaga}, M. and {Tegmark}, M. and {de Oliveira-Costa}, A.},
  title   = {E/B decomposition of finite pixelized CMB maps},
  journal = {Phys. Rev. D},
  volume  = {67},
  pages   = {023501},
  year    = {2003},
  doi     = {10.1103/PhysRevD.67.023501},
  eprint  = {astro-ph/0207338}
}

@article{Slosar2004,
  author  = {{Slosar}, A. and {Seljak}, U. and {Makarov}, A.},
  title   = {Exact likelihood evaluations and foreground marginalization in low resolution WMAP data},
  journal = {Phys. Rev. D},
  volume  = {69},
  pages   = {123003},
  year    = {2004},
  eprint  = {astro-ph/0403073}
}

@article{SmithZaldarriaga2007,
  author  = {{Smith}, K.~M. and {Zaldarriaga}, M.},
  title   = {General solution to the E-B mixing problem},
  journal = {Phys. Rev. D},
  volume  = {76},
  pages   = {043001},
  year    = {2007},
  eprint  = {astro-ph/0610059}
}

@article{Ferte2013,
  author  = {{Fert{\'e}}, A. and {Grain}, J. and {Tristram}, M. and {Stompor}, R.},
  title   = {Efficiency of pseudospectrum methods for estimation of the cosmic microwave background B-mode power spectrum},
  journal = {Phys. Rev. D},
  volume  = {88},
  pages   = {023524},
  year    = {2013},
  doi     = {10.1103/PhysRevD.88.023524},
  eprint  = {1305.7441}
}

@article{GarciaGarcia2019,
  author  = {{Garc{\'i}a-Garc{\'i}a}, C. and {Alonso}, D. and {Bellini}, E.},
  title   = {Disconnected pseudo-{$C_\ell$} covariances for projected large-scale structure data},
  journal = {JCAP},
  volume  = {11},
  pages   = {043},
  year    = {2019},
  doi     = {10.1088/1475-7516/2019/11/043},
  eprint  = {1906.11765}
}

@article{KnoxSong2002,
  author  = {{Knox}, L. and {Song}, Y.-S.},
  title   = {Limit on the detectability of the energy scale of inflation},
  journal = {Phys. Rev. Lett.},
  volume  = {89},
  pages   = {011303},
  year    = {2002},
  eprint  = {astro-ph/0202286}
}

@article{Kesden2002,
  author  = {{Kesden}, M. and {Cooray}, A. and {Kamionkowski}, M.},
  title   = {Separation of gravitational-wave and cosmic-shear contributions to cosmic microwave background polarization},
  journal = {Phys. Rev. Lett.},
  volume  = {89},
  pages   = {011304},
  year    = {2002},
  eprint  = {astro-ph/0202434}
}

@article{Pryke2009,
  author  = {{Pryke}, C. and others},
  collaboration = {QUaD},
  title   = {Second and Third Season QUaD Cosmic Microwave Background Temperature and Polarization Power Spectra},
  journal = {ApJ},
  volume  = {692},
  pages   = {1247},
  year    = {2009},
  doi     = {10.1088/0004-637X/692/2/1247},
  eprint  = {0805.1944}
}

@article{KamionkowskiKosowskyStebbins1997,
  author  = {{Kamionkowski}, M. and {Kosowsky}, A. and {Stebbins}, A.},
  title   = {A probe of primordial gravity waves and vorticity},
  journal = {Phys. Rev. Lett.},
  volume  = {78},
  pages   = {2058},
  year    = {1997},
  doi     = {10.1103/PhysRevLett.78.2058},
  eprint  = {astro-ph/9609132}
}

@article{SeljakZaldarriaga1997,
  author  = {{Seljak}, U. and {Zaldarriaga}, M.},
  title   = {Signature of gravity waves in the polarization of the microwave background},
  journal = {Phys. Rev. Lett.},
  volume  = {78},
  pages   = {2054},
  year    = {1997},
  doi     = {10.1103/PhysRevLett.78.2054},
  eprint  = {astro-ph/9609169}
}

@article{KamionkowskiKovetz2016,
  author  = {{Kamionkowski}, M. and {Kovetz}, E.~D.},
  title   = {The quest for {B} modes from inflationary gravitational waves},
  journal = {ARA\&A},
  volume  = {54},
  pages   = {227},
  year    = {2016},
  doi     = {10.1146/annurev-astro-081915-023433},
  eprint  = {1510.06042}
}

@article{ZaldarriagaSeljak1998,
  author  = {{Zaldarriaga}, M. and {Seljak}, U.},
  title   = {Gravitational lensing effect on cosmic microwave background polarization},
  journal = {Phys. Rev. D},
  volume  = {58},
  pages   = {023003},
  year    = {1998},
  doi     = {10.1103/PhysRevD.58.023003},
  eprint  = {astro-ph/9803150}
}

@article{HutererTakada2005,
  author  = {{Huterer}, D. and {Takada}, M.},
  title   = {Calibrating the nonlinear matter power spectrum: requirements for future weak lensing surveys},
  journal = {Astropart. Phys.},
  volume  = {23},
  pages   = {369},
  year    = {2005},
  doi     = {10.1016/j.astropartphys.2005.02.006},
  eprint  = {astro-ph/0412142}
}

@article{AmaraRefregier2008,
  author  = {{Amara}, A. and {R{\'e}fr{\'e}gier}, A.},
  title   = {Systematic bias in cosmic shear: extending the {Fisher} matrix},
  journal = {MNRAS},
  volume  = {391},
  pages   = {228},
  year    = {2008},
  doi     = {10.1111/j.1365-2966.2008.13880.x},
  eprint  = {0710.5171}
}

\begin{appendix}
\linenumbers    

\section{Construction of the filters}
\label{app:filters}

In this appendix, we give the construction of the filters of Sect.~\ref{sec:methods:filters}, all of them are matrices acting on the stacked $[Q\,|\,U]$ pixel vector of the footprint (Sect.~\ref{sec:methods:sims}).

\subsection{Isotropic filters}
\label{app:filters:isotropic}

The harmonic high-pass filter reads
\begin{equation}
  f_\ell \;=\; \sqrt{\,1 - \exp\!\left[-\left(\ell/\ell_0\right)^{\beta}\right]},
  \qquad \ell_0 = 30, \quad \beta = 6,
  \label{eq:softfilter}
\end{equation}
applied identically to $E$ and $B$. We build it as the pixel operator $\Rmat = \Imat + \mathbf{H}(f_\ell - 1)$, where $\mathbf{H}(g_\ell)$ applies to each footprint pixel the harmonic round trip of a spherical-harmonic analysis (\texttt{map2alm}), a multiplication by $g_\ell$, and a synthesis (\texttt{alm2map}), at the working band limit $\ell_{\rm sig}$. The eigenvalues of $\Rmat$ run from $6.8\times10^{-3}$ to $1$, the smallest set by how far a patch-restricted low multipole can be suppressed on the cut sky. With the rank threshold, only two directions are dropped (Appendix~\ref{app:checks} sweeps the threshold).

For the projector nulling the harmonics for $\ell < 30$, we render one beam-free template for each choice of $(\ell, m, \text{real/imaginary}, E\ \text{or}\ B)$, spanning $2 \le \ell \le 29$. Each is evaluated as a patch-restricted $QU$ map, obtaining $1792$ columns in total. Because most low-$\ell$ harmonics are near-degenerate over a patch this size ($\fsky = 0.10$), we orthonormalise them by an SVD and drop directions below $10^{-3}$ of the leading singular value. That leaves $514$ patch-supported modes, $21\%$ of the $QU$ rows, and the projector is $\Rmat = \Imat - \Proj$, with $\Proj$ the orthogonal projector onto the span of the retained left singular vectors.

\subsection{Stripe filters}
\label{app:filters:stripe}

The scan filter removes, from each constant-elevation stripe, the span of low-order polynomials in the scan coordinate,
\begin{equation}
  \begin{aligned}
  \Rmat &= \Imat - \Proj, \\
  \mathrm{range}(\Proj) &= \mathrm{span}\bigl\{\,
  P_k(x_s)\big|_{s,\,c}
  : k \le n,\ \forall s,\ c \in \{Q, U\} \bigr\},
  \end{aligned}
  \label{eq:scanfilter}
\end{equation}
where $\Proj$ is the orthogonal projector with that range, $P_k$ are the Legendre polynomials, $x_s$ is the longitude $\phi$ of the pixels in stripe $s$, rescaled to $[-1, 1]$ over the stripe, and the stripes are rings of constant colatitude $\theta$ about the \texttt{HEALPix} pole of the equatorial frame, that is, of constant declination. $Q$ and $U$ are filtered independently in the equatorial frame. The filter removes $172$, $336$ and $504$ modes at orders $n = 1, 3, 5$, against the projector's $514$.

The ground filter removes the sky structure constant in right ascension, since a template fixed in azimuth sweeps right ascension at constant declination as the sky rotates. On the sky, each template is therefore a function of declination alone, constant along every stripe and varying from one stripe to the next. The templates are labelled by the polynomial order $n$ they carry across the patch. The rank they remove is $2(n+1)$, before the singular values below $10^{-3}$ of the leading one are trimmed, which at orders 5, 10, 15, and 25 spans ranks from $12$ to $50$.

\subsection{Constant-elevation filters}
\label{app:site}

The constant-elevation filters are built from a toy scan strategy, checked against explicitly filtered time streams.

As regards the scan strategy, we place the telescope at latitude $-22.96^\circ$ and scan at a fixed elevation of $50^\circ$, sweeping $\pm 20^\circ$ in azimuth in $45\,$s. These choices follow the example SAT scanning strategy of \citet{SOtransfer}, for the SO site in the Atacama desert. Each field of the footprint is observed both rising and setting, from two fixed azimuth centres per field, in one-hour constant-elevation scans started every sidereal hour. The focal plane is represented by 21 effective detectors on a $7^\circ$ grid within a $35^\circ$ field of view, as in the hit map of Sect.~\ref{sec:methods:sims}, and the time streams are sampled every $0.5^\circ$ of azimuth. We assume an ideal rotating half-wave plate, so that each detector yields demodulated $Q$ and $U$ time streams in the frame of its own parallactic angle \citep{HWPmodulation}. That frame rotates against the sky along a sweep, which couples $Q$ and $U$ in the filtered map. In norm, the block of $\Rmat - \Imat$ that couples them is below a tenth of the block acting within $Q$.

The two filters follow the stripe filters of Appendix~\ref{app:filters:stripe}. The constant-elevation scan filter fits and subtracts a Legendre polynomial of order 3 in azimuth from each sweep of each detector, and the constant-elevation ground filter fits one of order 10 in azimuth over each whole scan of each detector, so that it captures signal fixed in azimuth. Each detector keeps a fixed elevation, at which an azimuth corresponds to a single declination. Like the stripe ground filter, it therefore removes structure constant in right ascension, here over the about $15^\circ$ the sky drifts during one scan. As in the main text, the two are applied separately. We bin the filtered time streams into a map with uniform weights, and the half-wave plate makes the hit counts $\mathbf{M}$ equal for $Q$ and $U$. Writing $\mathbf{A}$ for the pointing matrix, so that $\mathbf{M} = \mathbf{A}\tp\mathbf{A}$, and $\Proj_t$ for the projector onto the fitted templates in the time domain, the filter acting on the sky is then
\begin{equation}
  \Rmat = \mathbf{M}^{-1}\mathbf{A}\tp\left(\Imat - \Proj_t\right)\mathbf{A} .
  \label{eq:site_filter}
\end{equation}
Here, samples falling outside the footprint are left out of the fits, as for the stripe filters, while a real sweep also runs over the sky beyond it. A sweep or a scan that touches fewer than $n+1$ footprint pixels is left unfiltered. Explicit filtering and binning of simulated time streams reproduces Eq.~\eqref{eq:site_filter} to better than $10^{-10}$. The bottom row of Fig.~\ref{fig:maps_filters} shows the components the two filters remove from the $Q$ map.

For white time-stream noise of variance $\sigma^2$ per sample, the same map-maker gives the noise covariance
\begin{equation}
  \Nmat_{\Rmat} = \sigma^2\,\mathbf{M}^{-1}\mathbf{A}\tp\left(\Imat - \Proj_t\right)\mathbf{A}\,\mathbf{M}^{-1} = \sigma^2\,\Rmat\,\mathbf{M}^{-1} ,
  \label{eq:site_noise}
\end{equation}
since $(\Imat - \Proj_t)^2 = \Imat - \Proj_t$. This differs from the filtered covariance $\Rmat\Nmat\Rmat\tp$ of a filter defined in map space, as the stripe filters are. Indeed, in the basis in which $\mathbf{M}^{1/2}\Rmat\,\mathbf{M}^{-1/2}$ is diagonal, a mode attenuated by a factor $\gamma$ keeps a fraction $\gamma^2$ of its signal power and a fraction $\gamma$ of its noise power, so that its signal-to-noise ratio drops by $\gamma$, while under $\Rmat\Nmat\Rmat\tp$ it would be unchanged. We verified Eq.~\eqref{eq:site_noise} mode by mode in that basis, against the covariance of $2000$ noise realisations of one scan, filtered and binned explicitly. We set $\sigma^2$ so that the unfiltered map has the depth of Sect.~\ref{sec:methods:sims} at the mean hit count, and draw the noise realisations from Eq.~\eqref{eq:site_noise}.

\section{Structural checks on the restricted estimator}
\label{app:checks}

This appendix collects the checks on the restricted estimator listed in Sect.~\ref{sec:methods:qml}, together with the cost of conjugating only part of the QML algebra.

\subsection{Partially conjugated models}
\label{app:checks:partial}

\begin{figure*}[t]
  \centering
  \includegraphics[width=\textwidth]{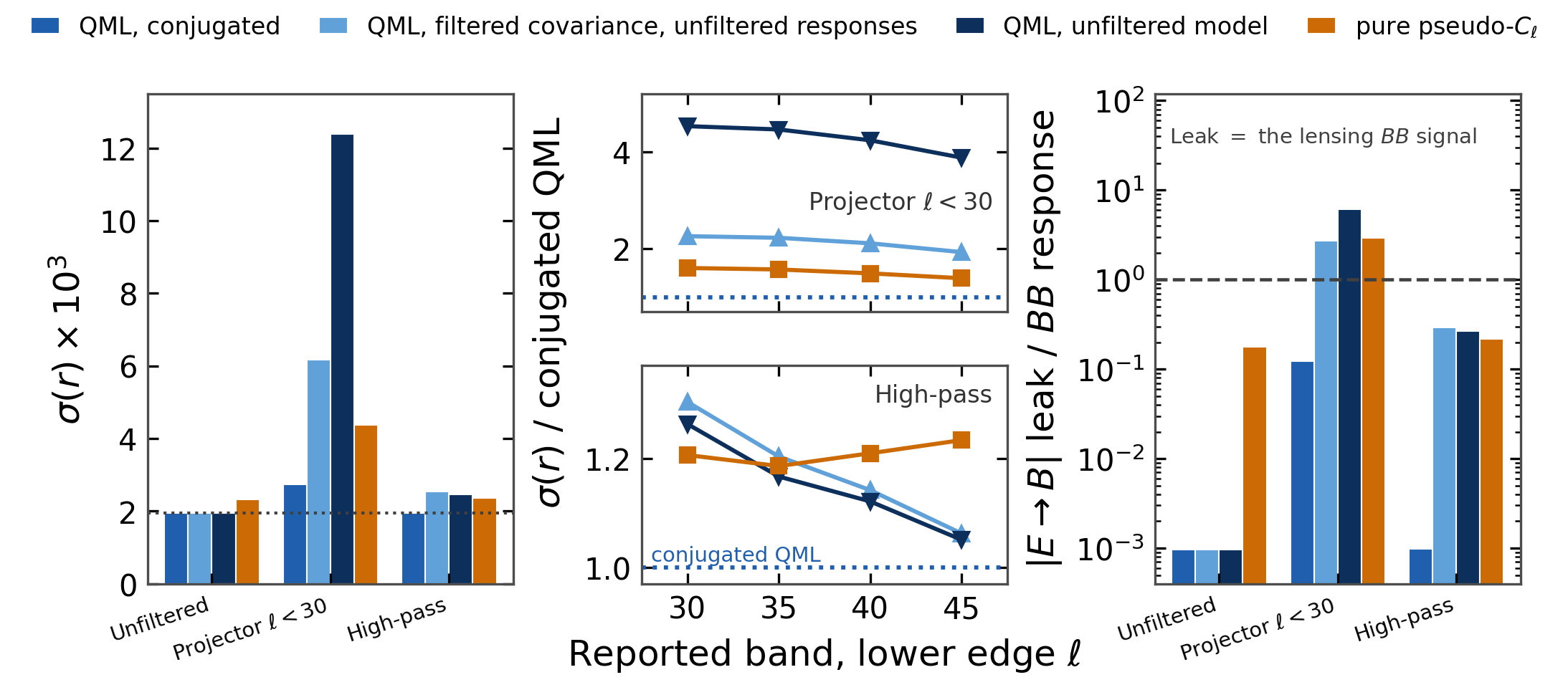}
  \caption{The four analysis routes at $2\ukam$, with bands buffered to $20 \le \ell \le 94$, on the first realisation set and with the tensor template extended over all multipoles. The pure \pcl pipeline is run at the same configuration as in Tables~\ref{tab:sigr_filtered} and~\ref{tab:leak}. Left: $\sigr$ per route over $30 \le \ell \le 64$, where the dotted line is the unfiltered QML. Middle: each route relative to conjugated QML against the lower edge of the reported band, under the projector (top) and the high-pass (bottom). Every curve stays above unity for every band under both filters, and under the projector the routes keep their full ordering throughout. Under the high-pass, \pcl crosses the unfiltered-model route once the band is narrowed to $\ell \ge 35$. Right: measured $|E \to B|$ leakage relative to the $BB$ lensing response, with the dashed line marking a leak equal to the lensing $BB$ signal. The leak is the summed ratio of Table~\ref{tab:leak}.}
  \label{fig:filtered_arms}
\end{figure*}

Figure~\ref{fig:filtered_arms} compares four analysis routes on the same filtered maps. The first is conjugated QML. The second runs QML with the covariance and the bandpower derivatives of an unfiltered sky (``unfiltered model''). The third conjugates only the noise covariance (``filtered covariance, unfiltered responses''). The fourth is pure \pcl. The middle panel of the figure shows each route against the lower edge of the reported band.

The partially conjugated route replaces the noise covariance with the filtered one, $\Rmat\Nmat\Rmat\tp$, plus a ridge of ten times the median pixel variance on the removed directions to keep it invertible, and leaves the signal covariance and the bandpower derivatives unconjugated. Under the projector, it costs $2.26\times$ at $2\ukam$ and $1.36\times$ at $3.5\ukam$ on $\sigr$, against $4.53\times$ and $3.12\times$ for the fully unconjugated model. Thus, filtering the noise model alone recovers about half of the cost in log terms at $2\ukam$, and three quarters at $3.5\ukam$. Since its derivatives are not conjugated, these numbers belong to that one ridge amplitude. Under the high-pass filter, the ordering inverts. There the partial model costs $1.30\times$ against $1.26\times$ at $2\ukam$, and $1.24\times$ against $1.17\times$ at $3.5\ukam$. Indeed, a graded filter scales the weight and the response by the same factor, so a model that is wrong in both largely cancels the error, while a model that corrects only the weight leans hardest on the modes it models worst. Note also that $\Rmat \Pmat_b \Rmat\tp = \Pmat_b$ holds only on the range of $\Rmat$, where conjugated QML is run. Both partial models run on the full pixel space, so the identity does not apply to their unconjugated derivatives.

\subsection{Exactness of the restriction}
\label{app:checks:restriction}

We report the checks of Sect.~\ref{sec:methods:qml}, run at $2\ukam$ on the first realisation set with bands from $\ell = 30$. First, the nulled subspace is affordable, since the projector removes $514$ of the $2456$ modes and four fifths of the pixel space is retained. Second, conjugation is exact for an invertible $\Rmat$: the hits weighting reproduces unfiltered QML to $6\times10^{-13}$. Third, the restriction matches a ridge-regularised conjugation, whose mean bandpowers agree with it to $3\times10^{-11}$ of their per-bin error over six orders of magnitude in the ridge amplitude, so nothing below depends on a regularisation choice (Fig.~\ref{fig:filtered_probe}, top). Fourth, a bin that the filter destroys shows a large variance rather than a quiet bias. In the $30 \le \ell \le 34$ bin, the $BB$ error grows by a factor of about $130$ and the mean bandpower runs at twice the unfiltered one, which is the signature of dividing by a vanished response. From $\ell = 40$ upward, the mean bandpowers agree with the unfiltered ones to within $5\%$, and the error inflation on $BB$ decays from $1.7$ in the $40 \le \ell \le 44$ bin to $1.07$ at the top of the band.

\subsection{The low-side buffer}
\label{app:checks:buffer}

\begin{figure}[t]
  \centering
  \includegraphics[width=\columnwidth]{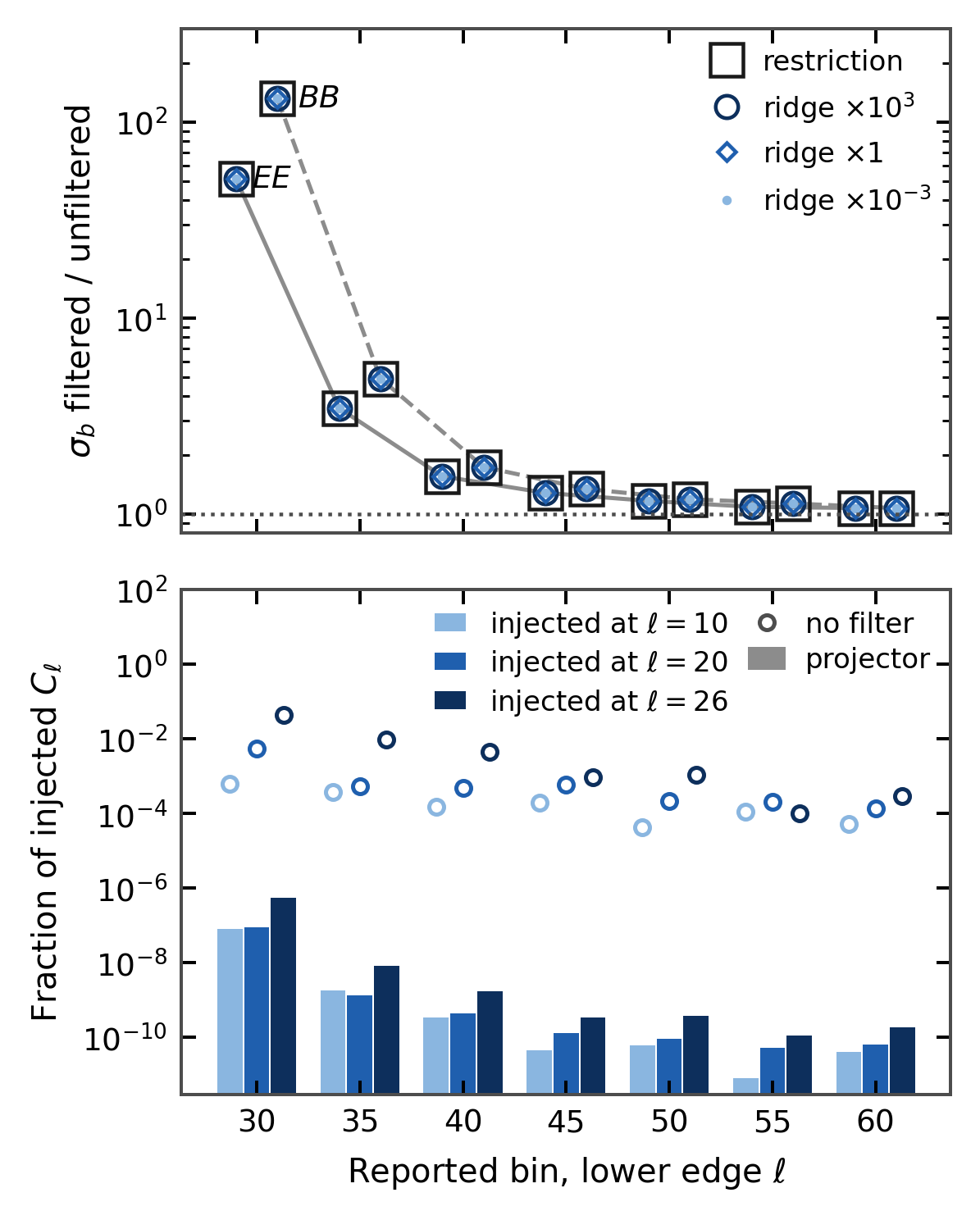}
  \caption{Top: $EE$ and $BB$ error inflation under the $\ell < 30$ projector in each reported bin, filtered against unfiltered, for the restriction of \qube (open squares) and for a ridge-regularised conjugation at three ridge amplitudes spanning six orders of magnitude. $EE$ sits left of each bin centre and $BB$ right of it. The four markers of each point coincide, so nothing downstream depends on the regularisation. Bottom: mean response in each reported bin to $BB$ power injected at $\ell = 10$, $20$ and $26$, below the reported band, in magnitude and as a fraction of the injected $C_\ell$, with no filter (open circles) and under the projector (bars). These runs carry no low-side buffer, so with no filter the injected power leaks into the reported bins through the bandpower window.}
  \label{fig:filtered_probe}
\end{figure}

Under the projector, $BB$ power injected below the reported band delivers essentially none of its response into the reported bins, suppressed by four to seven orders of magnitude (Fig.~\ref{fig:filtered_probe}, bottom). The projector therefore removes the need to model the low side, and no fiducial or foreground model is required there. The high-pass filter instead leaves $29\%$ of the amplitude at $\ell = 20$, by Eq.~\eqref{eq:softfilter}. Hence the low-side buffer is kept for both filters, which are then compared over the same band.

Under the projector, the Fisher matrix has a condition number of order $10^{17}$, at the limit of double precision. The cause is the $20 \le \ell \le 29$ buffer bins, which sit inside the nulled subspace. Indeed, rerunning the unfiltered and projector cases over $30 \le \ell \le 94$, without the low-side buffer, brings the condition number to $4.3\times10^{7}$, on the first realisation set. This moves QML's $\sigr$ by $1.3\%$ unfiltered and $2.1\%$ under the projector, and it moves the pure \pcl ratio in QML's favour by less than $0.1\%$ unfiltered and by $2.1\%$ under the projector. The ill-conditioning is confined to conjugated QML, since the two routes with unfiltered responses sit at $8.9\times10^{4}$. Thus, under the projector, the low-side buffer adds no information, only bins that conjugation then has to invert inside the null space.

\subsection{Rank and purification thresholds}
\label{app:checks:thresholds}

The high-pass filter is invertible, with condition number $150$, so by the argument of Sect.~\ref{sec:methods:qml} it destroys nothing. We nevertheless sweep the threshold below which a mode is discarded (Table~\ref{tab:rank_threshold}).
\begin{table}[t]
  \centering
  \small
  \caption{Rank-threshold sweep for the high-pass filter.}
  \label{tab:rank_threshold}
  \begin{tabular}{lcccc}
    \toprule
    Threshold & $10^{-1}$ & $10^{-2}$ & $10^{-3}$ & $\le 10^{-4}$ \\
    \midrule
    Modes kept & 2428 & 2454 & 2456 & $\ge 2456$ \\
    $\sigr/\sigr_{\rm unfilt}$ & 1.0039 & 1.0003 & 1.0000 & 1.0000 \\
    \bottomrule
  \end{tabular}
  \tablefoot{The pixel likelihood is evaluated in the reported window $30 \le \ell \le 64$. The threshold is the fraction of the leading singular value below which a mode is discarded, ``Modes kept'' is how many of the $2456$ pixel rows survive, and the last row is $\sigr$ relative to its unfiltered value.}
\end{table}
At the $10^{-2}$ threshold used for QML, two directions are dropped, and the cost is $0.03\%$; from $10^{-3}$ down, none are dropped. Thus, the rank threshold has no measurable effect on this filter.

The matrix purification of Eq.~\eqref{eq:purifier} has its own threshold, $\lambda_{\rm c}$, the smallest ratio of $B$ to $E$ power a kept direction may carry. We vary it over a decade around the value used in the main text (Table~\ref{tab:lambda_sweep}).
\begin{table}[t]
  \centering
  \footnotesize
  \caption{Sweep of the purification threshold $\lambda_{\rm c}$ of Eq.~\eqref{eq:purifier}.}
  \label{tab:lambda_sweep}
  \begin{tabular}{lccc}
    \toprule
    $\lambda_{\rm c}$ & 3 & 10 & 30 \\
    \midrule
    Kept, unfiltered / projector & 829 / 632 & 777 / 584 & 721 / 532 \\
    Purified bound, unfiltered & 1.010 & 1.015 & 1.023 \\
    Purified bound, projector & 1.031 & 1.046 & 1.068 \\
    Purified/QML, unfiltered & 1.003 & 1.006 & 1.014 \\
    Purified/QML, projector & 1.034 & 1.048 & 1.064 \\
    Leak, unfiltered ($10^{-4}$) & 6.0 & 2.5 & 1.2 \\
    \bottomrule
  \end{tabular}
  \tablefoot{Values are on the first realisation set. ``Kept'' counts the directions the purifier keeps, ``Purified bound'' is the pixel-likelihood bound on the purified data (Sect.~\ref{sec:methods:pics}) relative to the full one, ``Purified/QML'' is the ratio of $\sigr$, and the leak is defined as in Table~\ref{tab:leak}.}
\end{table}
Since the kept directions carry almost no $E$ power, $\lambda_{\rm c}$ and the ridge $\eta$ enter mainly through their product (Sect.~\ref{sec:methods:pcl}), so this sweep also covers the choice of $\eta$. A lower threshold keeps more directions and more of the information on $r$, at the price of more $E$ power left in the map, and the ratio to QML follows the information the purification discards. Over this range, the $\sigr$ ratio moves by at most $3\%$, and the leak stays two orders of magnitude below the pure \pcl one.

\subsection{Bin width and resolution}
\label{app:checks:binning}

We check that the comparison does not depend on the bin width by repeating it in bins of $\Delta\ell = 10$ in five states and on the first realisation set (Table~\ref{tab:binning}). The reported band becomes $30 \le \ell \le 69$, four bins, with the buffers at $20 \le \ell \le 29$ and $70 \le \ell \le 89$. The pixel-likelihood bound involves no binning and is unchanged. The pure \pcl search is repeated at the new bin width and returns the same configuration in every state.

We also repeat the comparison at $N_{\rm side} = 64$, in the same five states and on the first realisation set. The footprint is rebuilt from the hits map at the new resolution with the same sky fraction, and carries $9830$ $QU$ rows against $2456$. The bins, the reported band and the tensor template are those of the main analysis, while the working band limit follows the resolution to $\ell_{\rm sig} = 191 = 3N_{\rm side} - 1$. The simulations are drawn anew, so that the table also carries the scatter between realisation sets, which reaches $3\%$ on the pure \pcl ratios and $1.2\%$ on the matrix-purified ones (Table~\ref{tab:sigr_filtered}). To remove it, we analyse the same signal realisations at $N_{\rm side} = 32$ with $\ell_{\rm sig} = 95$, as in the main analysis, and, unfiltered and under the projector, at $N_{\rm side} = 64$ with $\ell_{\rm sig} = 95$, which separates the pixelisation from the band limit. The projector now removes $518$ modes, about the same as at $N_{\rm side} = 32$, since the subspace it nulls is set by the angular scale and the patch area, so that the share of the pixel space it removes falls from $21\%$ to $5\%$. For the scan filter, instead, we merge each pair of adjacent rings into one stripe, which keeps the stripe width of $N_{\rm side} = 32$. With one stripe per ring, the filter would remove $688$ modes, twice as many, while the width of a stripe is set by the sweep of the telescope, independently of the map resolution. At order 3, the merged stripes remove $352$ modes against $336$, the difference coming from the boundary of the rebuilt footprint. The high-pass filter and the ground template are defined independently of the pixelisation.
\begin{table*}[t]
  \centering \small \setlength{\tabcolsep}{4pt}
  \caption{Robustness to the bin width and the resolution.}
  \label{tab:binning}
  \begin{tabular}{lccccccccc}
    \toprule
    & \multicolumn{3}{c}{Baseline} & \multicolumn{3}{c}{$\Delta\ell = 10$} & \multicolumn{3}{c}{$N_{\rm side} = 64$} \\
    \cmidrule(lr){2-4}\cmidrule(lr){5-7}\cmidrule(lr){8-10}
    State & QML/bound & Pure/QML & Purified/QML & QML/bound & Pure/QML & Purified/QML & QML/bound & Pure/QML & Purified/QML \\
    \midrule
    Unfiltered & 1.033 & 1.178 & 1.006 & 1.031 & 1.166 & 1.009 & 1.007 & 1.102 & 1.000 \\
    High-pass  & 1.031 & 1.203 & 1.014 & 1.029 & 1.195 & 1.018 & 1.007 & 1.121 & 1.015 \\
    Projector  & 1.017 & 1.606 & 1.048 & 1.024 & 1.570 & 1.055 & 1.038 & 1.513 & 1.013 \\
    Scan 3     & 1.026 & 2.438 & 1.015 & 1.021 & 2.428 & 1.013 & 1.010 & 2.247 & 1.007 \\
    Ground 10  & 1.037 & 1.178 & 1.004 & 1.034 & 1.167 & 1.007 & 1.005 & 1.104 & 1.003 \\
    \bottomrule
  \end{tabular}
  \tablefoot{Values are on the first realisation set at $2\ukam$, with the tensor template confined to $30 \le \ell \le 64$ and the lensing amplitude fixed. The baseline is the configuration of Table~\ref{tab:sigr_filtered}, with bins of $\Delta\ell = 5$ at $N_{\rm side} = 32$. The middle group uses bins of $\Delta\ell = 10$, and the right group repeats the baseline at $N_{\rm side} = 64$. ``QML/bound'' is QML's $\sigr$ relative to the pixel-likelihood bound in the same state and configuration, while ``Pure/QML'' and ``Purified/QML'' give $\sigr$ of the pure and of the matrix-purified \pcl pipeline relative to QML's.}
\end{table*}
At $\Delta\ell = 10$, QML's $\sigr$ moves by at most $0.7\%$, while the pure \pcl one falls by $0.9$ to $1.6\%$, so the pure \pcl ratio falls by $0.4$ to $2.2\%$, most under the projector. The matrix-purified ratio moves by at most $0.7\%$. Thus, neither the pure \pcl gap nor the distance of QML from the bound comes from the narrow bins.

At $N_{\rm side} = 64$, the pixel-likelihood bound on the purified data falls by $1.6$ to $3.0\%$ in every state, most under the projector. The cost of each filter on that bound, relative to the unfiltered state, moves by at most $1.5\%$. On the same signal realisations, QML's $\sigr$ moves by at most $1.7\%$ and the matrix-purified ratio falls by at most $2.2\%$, while the pure \pcl ratio falls by $5.9$ to $11.4\%$, least under the projector. The change is in the pure \pcl $\sigr$ itself, and it comes from the pixelisation rather than from the band limit. Indeed, at $N_{\rm side} = 64$ with $\ell_{\rm sig} = 95$ the pure \pcl $\sigr$ is lower than at $N_{\rm side} = 32$ by $10.6\%$ unfiltered and by $12.2\%$ under the projector, and raising the band limit to $\ell_{\rm sig} = 191$ then increases it by $2.2$ and $6.5\%$. With the finer pixels, the configuration search also selects a narrower taper in every state but the scan filter, $7.5^\circ$ instead of $10^\circ$ unfiltered and $10^\circ$ instead of $15^\circ$ under the projector. At a fixed taper, the finer pixels already lower the pure \pcl $\sigr$ by $5.1\%$ unfiltered and by $3.1\%$ under the projector, about $60$ and $50\%$ of the change, while at $N_{\rm side} = 32$ the narrower taper would raise it by $16\%$ unfiltered. Over the three realisation sets, the pure \pcl ratio at $N_{\rm side} = 64$ is $1.104 \pm 0.007$ unfiltered and $1.46 \pm 0.04$ under the projector, against $1.21$ and $1.56$ at $N_{\rm side} = 32$ (Table~\ref{tab:sigr_filtered}), and the matrix-purified one is $1.005$ and $1.02$. Thus, at $N_{\rm side} = 32$, about half of the unfiltered pure \pcl gap comes from the coarse pixelisation, while the ordering of the filters and the agreement of the matrix-purified pipeline with QML hold at both resolutions.

On one node with 48 cores, a single QML computation with the default computation basis of \qube, that is building the Fisher matrix and estimating the bandpowers of one map, takes about $50\,$s at $N_{\rm side} = 32$ and about $4\,$min at $N_{\rm side} = 64$, where \qube selects its harmonic computation basis instead of the pixel one. Estimating the other 999 maps of a set adds only a few seconds, since nearly all the cost is in the Fisher matrix and in the setup of the estimator.

\section{The unfiltered \pcl search and the analytic covariances}
\label{app:unfiltered}

This appendix collects the two parts of the unfiltered comparison that involve no filter. The first is the \pcl configuration search, which selects the pure \pcl setup and justifies the reduced grid of Sect.~\ref{sec:methods:pcl}. The second is the fidelity of the analytic covariances of both estimators. Both are measured over $30 \le \ell \le 64$ alone, with no buffer bins.

\subsection{The \pcl configuration search}
\label{app:unfiltered:search}

The first question is how well a tuned \pcl can do, and how much the tuning matters. To answer it, Table~\ref{tab:sat_sweep} collects the ratio of amplitude errors $\sigA(\text{\pcl})/\sigA(\mathrm{QML})$ over $30 \le \ell \le 64$ for the 78 configurations of the search. Here, $\sigA$ is Eq.~\eqref{eq:sigmaA} with $\mathbf{t}_0$ replaced by the measured response to the lensed $EE$ or $BB$ spectrum, and $\boldsymbol{\Sigma}$ by the covariance of that spectrum's bandpowers, since $EE$ has no parameter analogue of $r$. A ratio of one means that the two estimators are equally good, while a ratio above one favours QML. Note that the $BB$ entries use the lensing template rather than the tensor one, although the two give pure \pcl ratios that agree to within $5\%$ in every filter state\footnote{The two templates weight the band differently, since over $30 \le \ell \le 64$ the lensing $BB$ spectrum is nearly white while the tensor one falls by almost $40\%$.}. For these reasons, the tuned $\rho_{BB}$ differs slightly from the $\sigr$ ratio of Sect.~\ref{sec:results}. The bands are not buffered, and the search runs on the first of the three realisation sets of Sect.~\ref{sec:methods:sims}. We find that a tuned \pcl trails QML by about a fifth on $BB$ and by a quarter to a third on $EE$, while an untuned one could be several times worse. The matrix-purified pipeline is not part of this search, which also scores $EE$, since its maps carry almost no $E$ power. Its own grid is described in Sect.~\ref{sec:methods:pcl}.

\begin{table}[t]
  \centering \footnotesize \setlength{\tabcolsep}{3pt}
  \caption{The unfiltered configuration sweep.}
  \label{tab:sat_sweep}
  \begin{tabular}{llccc}
    \toprule
    & Configuration & $\rho_{EE}$ & $\rho_{BB}$ & Leak \\
    \midrule
    Binary baseline    & Binary, unpurified, uniform              & 1.08 & 12.99 & 4.50 \\
    \midrule
    Best joint         & $10^\circ$ C2, $EB$ pur., uniform        & 1.27 & 1.22 & 0.24 \\
    Best $BB$          & $10^\circ$ C2, $B$ pur., uniform         & 1.34 & 1.22 & 0.24 \\
    Lowest leak at $10^\circ$ & $10^\circ$ C2, $B$ pur., hits     & 1.47 & 1.22 & 0.14 \\
    \midrule
    Apodisation only   & $10^\circ$ C2, unpurified, uniform       & 1.34 & 7.44 & 0.85 \\
    Purification only  & Binary, $B$ pur., uniform                & 1.08 & 11.96 & 15.05 \\
    \bottomrule
  \end{tabular}
  \tablefoot{The band is $30 \le \ell \le 64$ with $\Delta\ell = 5$, and the 78 configurations are run over 1000 realisations of the first set at $30'$ and $2\ukam$. Here $\rho_X \equiv \sigA(\text{\pcl})/\sigA(\mathrm{QML})$ for spectrum $X$, and ``Leak'' is the measured $E \to B$ contamination in units of the $BB$ signal (Sect.~\ref{sec:methods:protocol}). The joint and $BB$ optima share a mask and differ only in whether $E$ is purified, which cannot move a $BB$ number. The lower block isolates each remedy, and with both, the configuration is the best joint one.}
\end{table}

\subsection{Apodisation, purification and weighting}
\label{app:unfiltered:mask}

Table~\ref{tab:sat_sweep} also shows where this cost comes from. On an untreated binary mask, the $BB$ error is more than ten times QML's. Indeed, the sharp edge of the mask lets $E$ power leak into $B$, and since $E$ is much larger than $B$, even a small leak dominates the $BB$ scatter. The two standard remedies only work as a pair, as expected. Purification removes the leak generated by the mask, but it needs to differentiate the mask, so it only works on a smooth one. Apodisation provides that smooth mask, but on its own it only reduces the leak without removing it. With both on, the leak keeps falling as the taper widens, long after the error bar has started to worsen, down to a tenth of the $BB$ signal at the widest hit-weighted taper. These leaks are measured on unbuffered bands, and are somewhat larger than those of Table~\ref{tab:leak}, which uses the buffered bands of Sect.~\ref{sec:methods:qml}. The best compromise is a $10^\circ$ taper, about five pixels at $N_{\rm side} = 32$, and $BB$ degrades on both sides of it. A narrower taper is too sharp for purification, while a wider one down-weights more of the patch, and so uses less of the data. Even the shape of the taper matters, since the \texttt{C2} kernel beats \texttt{C1} at every taper in the grid.

Furthermore, $EE$ and $BB$ prefer incompatible masks. The leak from $B$ into $E$ is negligible against the $EE$ signal, so $EE$ has nothing to gain from purification, and it is best on the untreated binary mask, which keeps every pixel at full weight. Every treatment that helps $BB$ makes $EE$ worse. Thus, any single \pcl analysis must compromise between the two spectra. On the other hand, QML estimates both $EE$ and $BB$ at once, from one likelihood, with no compromise and no tuning.

Hit weighting is the remaining knob. On an untreated mask, it helps about as much as apodisation does. This is because the hit count falls towards the edge of the footprint, so that weighting the pixels by it acts as a taper. At the tuned optimum, where the mask is already tapered, it buys nothing on $BB$ and costs $EE$, and what it buys there is a lower leak (Table~\ref{tab:sat_sweep}). In fact, on $\sigr$ the best pure \pcl configuration in every unfiltered case uses uniform weighting, with the best hit-weighted one close behind, as expected if the weighting substitutes for a taper rather than adding to one. Thus, what remains at the pure \pcl optimum is not something that pixel weighting can remove. Section~\ref{sec:results:ratio} shows that it is the cost of separating $E$ from $B$ harmonically, together with the apodisation that harmonic purification requires.

\subsection{Analytic covariances}
\label{app:unfiltered:cov}

\begin{table}[t]
  \centering \footnotesize \setlength{\tabcolsep}{4pt}
  \caption{Fidelity of the analytic error bars.}
  \label{tab:cov_fidelity}
  \begin{tabular}{lcccccc}
    \toprule
    & \multicolumn{2}{c}{QML, $\Fmat^{-1}$} & \multicolumn{2}{c}{\texttt{NaMaster}, tuned} & \multicolumn{2}{c}{\texttt{NaMaster}, binary} \\
    \cmidrule(lr){2-3}\cmidrule(lr){4-5}\cmidrule(lr){6-7}
    $\ell$ bin & $EE$ & $BB$ & $EE$ & $BB$ & $EE$ & $BB$ \\
    \midrule
    30--34 & 0.99 & 1.04 & 1.05 & 1.13 & 0.98 & 1.13 \\
    35--39 & 1.00 & 1.01 & 1.00 & 1.05 & 0.97 & 1.16 \\
    40--44 & 0.98 & 0.98 & 1.03 & 1.02 & 0.96 & 1.10 \\
    45--49 & 1.01 & 1.01 & 1.02 & 1.03 & 0.98 & 1.07 \\
    50--54 & 1.01 & 0.99 & 1.02 & 1.01 & 0.98 & 1.10 \\
    55--59 & 0.95 & 0.98 & 0.99 & 0.98 & 0.94 & 1.08 \\
    60--64 & 1.00 & 0.98 & 0.97 & 1.02 & 0.97 & 0.99 \\
    \bottomrule
  \end{tabular}
  \tablefoot{Each entry is the ratio of the analytic to the empirical bandpower standard deviation in that reported bin, over the 1000 realisations of the first set at $2\ukam$. QML's analytic error is $\Fmat^{-1}$ of Eq.~\eqref{eq:qml_fisher}, \texttt{NaMaster}'s is its Gaussian bandpower covariance in the improved narrow-kernel approximation \citep{Nicola2021}, fed the mean measured pseudo-spectra, evaluated at the tuned configuration and on the untreated binary mask. The Monte Carlo precision of such a ratio is about $2\%$.}
\end{table}

A practical question is whether the error bars of each estimator can be trusted without simulations. Both estimators come with an analytic bandpower covariance, and Table~\ref{tab:cov_fidelity} compares each with its own empirical scatter. For QML, $\Fmat^{-1}$ of Eq.~\eqref{eq:qml_fisher} matches the empirical errors in every bin of both spectra, at the Monte Carlo precision of a 1000-realisation covariance. This is expected, since the covariance model of QML describes these simulations exactly (Sect.~\ref{sec:methods:sims}). \texttt{NaMaster}'s Gaussian bandpower covariance is nearly as accurate at the tuned configuration, provided it is fed the measured pseudo-spectra divided by the mean squared weight. This is the improved narrow-kernel approximation of \citet{Nicola2021}, which builds on \citet{GarciaGarcia2019}. Its largest deviations are in the lowest $BB$ bin and on the untreated binary mask. If it is fed the theory spectra instead, it underestimates the error by a factor of two to three on $BB$ on the binary mask. This is because the leaked $E$ power that dominates the untreated $BB$ scatter is present in the measured pseudo-spectra, but not in the theory input. Thus, with the input it is designed for, neither estimator needs simulations for its error bars on this patch.

\section{The matrix transfer function as a bias correction}
\label{app:tf}

This appendix describes what the matrix transfer function of Eq.~\eqref{eq:tf} does to the mean of the pure \pcl bandpowers, compares the construction with published ones, and measures its effect on the $\sigr$ of this work.

\subsection{The mean leak and the bias on $r$}
\label{app:tf:mean}

We measure the matrix of Sect.~\ref{sec:methods:pcl} on the first realisation set, at the pure \pcl configuration of Table~\ref{tab:leak}. We then score the residual $E \to B$ leak against the $BB$ lensing response over $30 \le \ell \le 64$, as in Table~\ref{tab:leak}. Unfiltered and under the high-pass filter, the correction removes $92\%$ and $85\%$ of the leak, and the residuals, $0.015 \pm 0.001$ and $0.033 \pm 0.001$, are more than ten times their Monte Carlo errors. Under the scan filter, it removes $96\%$ of the leak at order 1, where the residual is $0.018 \pm 0.004$, and more than $99\%$ at orders 3 and 5, where the residuals, $0.006 \pm 0.005$ and $0.007 \pm 0.007$, are consistent with zero. Under the constant-elevation scan filter, the residual is $0.160 \pm 0.051$. The matrix has no column below $\ell = 20$ or above $\ell = 94$, so power there enters the reported bins without being modelled. Most of the residual is $E$ power below $\ell = 20$. Adding columns down to $\ell = 2$ lowers the corrected leak to $0.008 \pm 0.001$ unfiltered and $0.012 \pm 0.001$ under the high-pass filter.

Under the projector, the full $2n_b \times 2n_b$ system has a condition number of $1.2\times10^{12}$, against $8$ to $20$ in the other states, since its $20 \le \ell \le 29$ columns probe the subspace the projector nulls. This is the same cause as the ill-conditioning of QML's buffer bins (Appendix~\ref{app:checks:buffer}). We therefore build it on the $\ell \ge 30$ columns only and solve for the kept bins by least squares. Its condition number is then $5.0\times10^{3}$, and the corrected leak, $0.6 \pm 6.0$, is dominated by Monte Carlo noise, since the inversion amplifies the scatter of the $E$-only bandpowers.

The bias the mean leak puts on $r$ (Sect.~\ref{sec:methods:protocol}) does not suffer from this, since it weights the bins through the covariance that the inversion amplifies in the same way. We measure it on the same coupled bandpowers, before and after $\Tmat^{-1}$. Before the correction, it is $1.8$ to $2.7\,\sigr$ in every state. After it, it is $0.10$, $0.13$ and $0.07\,\sigr$ unfiltered, under the high-pass filter and at scan order 1, and below $0.03\,\sigr$ at scan orders 3 and 5 and under the projector. With the columns down to $\ell = 2$, it falls below $0.04\,\sigr$ unfiltered and under the high-pass filter. A matrix transfer function therefore repairs the mean in every state, provided it leaves out the scales the filter nulls.

\subsection{Comparison with published constructions}
\label{app:tf:published}

Published filter corrections differ in what they couple. The Atacama B-mode Search (ABS) \citep{HWPmodulation} divides each bandpower by one scalar transfer function, the same for $EE$, $BB$ and $EB$, so its filter correction couples neither spectra nor bandpowers. The validation of filtered SAT spectra by \citet{SOtransfer} uses a $4 \times 4$ matrix per bandpower, over $EE$, $EB$, $BE$ and $BB$, fitted to filtered power-law simulations. It couples the spectra within a bandpower, while different bandpowers are coupled only through the mask. Our $\Tmat$ couples both, over $[EE\,|\,BB]$ and across all bandpowers, as the per-multipole bandpower window functions of BICEP/Keck \citep{TODfiltering, ObsMatrix} do, which \citet{SOtransfer} use as their reference. We do not carry the $EB$ and $BE$ columns, and this is exact in the mean. \citet{SOtransfer} carry them to capture the mixing between polarisation channels by the filter. Our $Q$ and $U$ maps are generated from $E$ and $B$ harmonic coefficients drawn independently (Sect.~\ref{sec:methods:sims}), so the true $C^{EB}_\ell$ on which $\Tmat$ acts is zero, the $EB$ and $BE$ columns multiply zero, and we do not use the $EB$ output rows.

\subsection{Effect on $\sigr$}
\label{app:tf:sigr}

The $\sigr$ of this work is computed on the reported $BB$ bins only, while $\Tmat^{-1}$ mixes $EE$ and buffered bins into them. We therefore apply $\Tmat^{-1}$ to the coupled $[EE\,|\,BB]$ bandpowers of the signal-plus-noise, tensor, and lensing sets at the same configurations, and score $\sigr$ on the reported $BB$ bins before and after. The correction changes $\sigr$ by $+1.3\%$ unfiltered, $+0.8\%$ under the high-pass filter, $+1.0\%$, $+1.4\%$ and $-0.2\%$ under the scan filter at orders 1, 3 and 5, and $+5.6\%$ under the projector, against pure \pcl gaps of $21$ to $204\%$. Relative to the decoupled bandpowers at the same configurations, the corrected $\sigr$ is $0.1$ to $0.7\%$ larger, and $1.3\%$ under the projector. Thus, the transfer function removes the mean leak without closing the gap.

\section{The pixel likelihood in the signal-to-noise eigenbasis}
\label{app:pencil}

The covariance of Eq.~\eqref{eq:pics_like} splits exactly into an $r=0$ part and an $r=1$ tensor template scaled by $r$, since the tensor spectrum is proportional to $r$ and the scalar sky does not depend on it. This appendix derives from that split every property of the pixel likelihood used in the main text. The key step is a change of basis to the signal-to-noise eigenmodes of the tensor template, the Karhunen--Lo\`eve basis of \citet{TTH1997}, in which the likelihood becomes a sum over independent modes.

\subsection{The statistic}
\label{app:pencil:statistic}

Both matrices in Eq.~\eqref{eq:pics_like} are built from the same inputs as the simulations (Sect.~\ref{sec:methods:sims}). This matters more here than for the bandpower estimators. Indeed, with no bandpowers there is no buffer, so the likelihood sees every mode up to $\ell_{\rm sig}$, and any mismatch between the modelled and the simulated truncation would appear as spurious information. $\Smat_{\rm t}$ contains only the tensor $BB$ spectrum. Following Paper~I, we report the mean-$\chi^2$ likelihood $\bar{\mathcal{L}}(r) \propto \exp[-\tfrac12\langle\chi^2(r)\rangle_{\rm sims}]$, with $\chi^2(r) \equiv -2\ln\mathcal{L}(r)$ of Eq.~\eqref{eq:pics_like}. It peaks at $r_{\rm true}$ for a correct covariance model, so that the position of its peak is itself a test of unbiasedness. Since $r \ge 0$ is a boundary and $r_{\rm true} = 0$, we quote the curvature as the Fisher-comparable number, and one-sided $68\%$ and $95\%$ credible upper limits as the posterior statement.

\subsection{Restriction, and the determinant a projector destroys}
\label{app:pencil:restrict}

The quadratic form in Eq.~\eqref{eq:pics_like} conjugates as in Eq.~\eqref{eq:conjugation}, but $\ln|\Cmat|$ does not. For a projector, $\Rmat \Cmat \Rmat\tp$ is exactly singular and its log-determinant is undefined. No regularisation is needed to resolve this. Since $\Rmat = \Imat - \Proj$ acts as the identity on its own range, $\Rmat \Umat = \Umat$ for any $\Umat$ spanning the range of $\Rmat$, so
\begin{equation}
  \Umat\tp\!\left(\Rmat \Cmat \Rmat\tp\right)\!\Umat = \Umat\tp \Cmat \Umat,
  \qquad
  \Umat\tp\!\left(\Rmat \dvec\right) = \Umat\tp \dvec ,
  \label{eq:pics_restrict}
\end{equation}
and the likelihood of the filtered data is exactly the likelihood on the surviving subspace. Each state is then described by a single matrix. It is the identity when unfiltered and the diagonal hits weighting for the invertible control, while for the other filters it is the rectangular $\Umat\tp\Rmat$ of Sect.~\ref{sec:methods:qml}, which for the projector, the scan filters and the ground template reduces to $\Umat\tp$, with $2456$ rows less the ``Modes'' of Table~\ref{tab:sigr_filtered}.

\subsection{Two transformations and the closed form}
\label{app:pencil:closed}

Write $\Cmat_{\Rmat}(r) = \Cmat_0 + r\,\Cmat_{\rm t}$ with $\Cmat_0 = \Rmat(\Smat_{\rm s} + \Nmat)\Rmat\tp$ the $r=0$ model and $\Cmat_{\rm t} = \Rmat \Smat_{\rm t} \Rmat\tp$, and Cholesky-factor the first, $\Cmat_0 = \Lchol\Lchol\tp$.

The first transformation whitens the data: set $\xvec = \Lchol^{-1}\Rmat\dvec$ and under the $r=0$ model $\langle \xvec \xvec\tp\rangle = \Imat$, so a correct covariance model gives white data.

The second rotates into the signal-to-noise eigenbasis of the tensor template. The matrix $\Lchol^{-1}\Cmat_{\rm t}\Lchol^{-\mathsf T}$ is symmetric, so it diagonalises as $\Omat\,\mathrm{diag}(\kappa)\,\Omat\tp$ with $\Omat$ orthogonal. Each eigenvalue $\kappa_i$ is the tensor variance of mode $i$ at $r=1$ in units of its variance at $r=0$, as in the Karhunen--Lo\`eve decomposition of \citet{TTH1997}. Thus, using $\yvec = \Omat\tp\xvec$ means that for every $r$ at once,
\begin{equation}
  \left\langle \yvec \yvec\tp \right\rangle
  = \Omat\tp\Lchol^{-1}\,\Cmat_{\Rmat}(r)\,\Lchol^{-\mathsf T}\Omat
  = \Imat + r\,\mathrm{diag}(\kappa) ,
  \label{eq:pencil_diag}
\end{equation}
which is diagonal, so that mode $i$ has variance $1 + r\kappa_i$ and is independent of the rest. Equivalently $\Cmat_{\Rmat}(r) = \Lchol\,\Omat\left(\Imat + r\,\mathrm{diag}(\kappa)\right) \Omat\tp\Lchol\tp$, so both terms of Eq.~\eqref{eq:pics_like} become sums over modes. Averaging $y_i^2$ over the simulation stack as $a_i = \langle y_i^2\rangle$ gives the whole posterior as a one-dimensional sum,
\begin{equation}
  -2\ln\bar{\mathcal{L}}(r) = \sum_i\left[
    \frac{a_i}{1 + r\kappa_i} + \ln\left(1 + r\kappa_i\right)\right]
  + \ln|\Cmat_0| .
  \label{eq:pics_pencil}
\end{equation}
After one eigendecomposition, each value of $r$ on the grid costs $\mathcal{O}(n_{\rm R})$. The posterior is therefore independent of the grid resolution by construction, and the ten states, over three realisation sets, cost about two minutes on one node.

The curvature at $r=0$ follows immediately and is analytic,
\begin{equation}
  \sigr = \left(\tfrac12\textstyle\sum_i \kappa_i^2\right)^{-1/2} ,
  \label{eq:pics_sigma}
\end{equation}
containing no simulated covariance and therefore carrying no Monte Carlo scatter, unlike the QML and \pcl columns of Table~\ref{tab:sigr_filtered}.

\subsection{Controls}
\label{app:pencil:controls}

We run three controls, and the reported numbers pass all of them.

First, an invertible $\Rmat$ cannot change the likelihood. Indeed, the diagonal hits weighting reproduces the unfiltered $\sigr$ to $9\times10^{-16}$, as the invertibility test of Sect.~\ref{sec:methods:qml} does for QML.

Second, we check the restriction against a ridge. Evaluating Eq.~\eqref{eq:pics_like} by a ridge-regularised conjugation instead agrees with the exact restriction of Eq.~\eqref{eq:pics_restrict} to $10^{-15}$, across four orders of magnitude in the ridge amplitude, under the projector and the order-3 scan filter.

Third, we test the whitening. The test needs nothing beyond the $a_i$ of Eq.~\eqref{eq:pics_pencil}, which the posterior already uses. Since $r_{\rm true} = 0$, Eq.~\eqref{eq:pencil_diag} gives $\langle y_i^2\rangle = 1$ in every mode for a correct model. Each $a_i$ is a $\chi^2_1$ variable averaged over $n_{\rm sim}$ realisations, so its expectation is $1 \pm \sqrt{2/n_{\rm sim}} = 1 \pm 0.045$. Averaged over modes, we measure $\bar a$ between $0.998$ and $1.000$ in every filter state and realisation set.

This test matters because of what a failure does to the posterior. Differentiating Eq.~\eqref{eq:pics_pencil} at $r=0$ gives
\begin{equation}
  \left.\frac{\partial\left(-2\ln\bar{\mathcal{L}}\right)}{\partial r}
  \right|_{r=0} = -\sum_i \kappa_i \left(a_i - 1\right) ,
  \label{eq:pencil_grad}
\end{equation}
Thus, with every $a_i = 1$, the peak sits exactly at the true value, $r=0$. A model covariance that underestimates the variance in some direction leaves the simulations with more power there than expected, $a_i > 1$, and moves the peak to positive $r$, so that a covariance error is read as a tensor detection. This is why the test is run in this basis. Any orthogonal rotation of a whitened vector has unit variance mode by mode, so the mean test passes in every basis. Only in this one, however, is each deviation weighted by the tensor response $\kappa_i$ of its direction.

Note that this is the only control in this work that compares the simulations with the full model covariance. The comparison in Appendix~\ref{app:unfiltered:cov} tests the covariance one bandpower at a time, while this test covers all $2456$ directions at once, at no cost beyond the $a_i$ the posterior already needs.

With a finite set of $n_{\rm sim} = 1000$ realisations, the peak wanders around the $r=0$ edge by $1/\sqrt{n_{\rm sim}}$ in units of $\sigr$, independently of the state. Over all states and realisation sets, we measure at most $0.014\,\sigr$, within the expected scatter of $0.032\,\sigr$.

\end{appendix}
\end{document}